\documentclass[acmsmall,screen, nonacm]{acmart}
\usepackage{amsfonts}
\usepackage{textcomp}
\usepackage[ruled,vlined,linesnumbered]{algorithm2e}
\usepackage{pifont}

\AtBeginDocument{%
  }

\SetKwProg{Fn}{function}{}{end}
\SetKwFunction{FEnrollTables}{Enroll\_Tables}
\SetKwFunction{FWindowEncode}{Window\_Encode}
\SetKwFunction{FRestore}{Restore}
\SetKwFunction{FRestored}{Restored}
\SetKwFunction{FMask}{Mask}
\SetKwFunction{FNextAddr}{NextAddr}
\SetKwFunction{FTryVisit}{TryVisit}
\SetKwFunction{FEnrollCRP}{Enroll\_CRP}
\SetKwFunction{FDUFT}{DUFT}
\SetKwFunction{FMSRAM}{MSRAM}
\SetKwFunction{FPowerCycle}{PowerCycle}
\SetKwFunction{FRead}{Read}
\SetKwFunction{FBCHCorrect}{BCH\_Correct}
\SetKwFunction{FBLAKE}{BLAKE2s}
\SetKwFunction{FtoBitArray}{toBitArray}

\setcopyright{none}
\begin{document}

\title{DUFT: Device-Unique Feedback Traversal for Entropy-Estimated Response Extraction and Mutual Authentication using SRAM PUFs}

\author{Syed Talha Imam}
\affiliation{%
  \institution{Department of Electrical Engineering and Computer Science, Florida Atlantic University}
  \city{Boca Raton}
  \state{Florida}
  \country{USA}
}

\author{Amna Zafar}
\affiliation{%
  \institution{Department of Electrical Engineering and Computer Science, Florida Atlantic University}
  \city{Boca Raton}
  \state{Florida}
  \country{USA}
}

\author{Arslan Munir}
\correspondingauthor
\affiliation{%
  \institution{Department of Electrical Engineering and Computer Science, Florida Atlantic University}
  \city{Boca Raton}
  \state{Florida}
  \country{USA}
}

\begin{abstract}
Static Random Access Memory (SRAM) Physical Unclonable Functions (PUFs) are a promising technology in the realm of random number generation for key generation, but for PUF-based authentication, they have been sidelined due to their limited challenge space as a reservoir of Challenge-Response Pairs (CRPs). 
In this paper, we develop a seed-based, round-iterated feedback mechanism, which we name Device-Unique Feedback Traversal (DUFT).
The feedback chain is driven by the device's unstable and stable bit positions in its stability mask against the address to create a unique traversal.
Additionally, a device-key hashed from the unique SRAM-PUF array is also included in calculating the next round of the chain to keep it secret. 
By estimating the physical SRAM source's entropy and error rate empirically, or by having a source with established metrics in both, DUFT estimates the number of rounds of feedback in the traversal to be performed to achieve a certain amount of min-entropy.
We reach 128 bits of estimated min-entropy on our reference hardware under stated assumptions and experimental estimates.
The number of rounds in the traversal scales directly with measured source-entropy, allowing security to be increased for weaker sources at the cost of running time.
Using this feedback-traversal wrapper over SRAM-PUFs to obtain a measured-response, we propose an extremely lightweight authentication and key-encapsulation protocol for resource-constrained embedded devices.
We realize and evaluate DUFT end-to-end on physical IoT/embedded hardware: a discrete-SRAM node pair running the complete mechanism and end-to-end protocol, and an embedded-SRAM microcontroller for the mechanism alone.
\end{abstract}


\keywords{Embedded Systems, Entropy, Feedback, Hardware Security, Internet of Things, Lightweight Cryptography, Mutual Authentication, Physical Unclonable Functions,  Resource-Constrained Devices, Static random-access memory}


\maketitle

\section{Introduction}
\label{section:introduction}
Physical Unclonable Functions (PUFs) gained popularity as an alternative to the public-key infrastructure and asymmetric key cryptography mechanisms to authenticate parties and exchange keys between them before starting secure communication with symmetric key cryptography~\cite{suh2007physical}. 
PUFs found their use in the realm of embedded computing and in the Internet of Things (IoT) space because of their self-occurring nature as the random and unique process variations inherent in physical devices. 
PUFs utilize the in-built inter-process variations in semiconductor devices, such as the signal delay of a wire~\cite{gassend2002silicon} or the startup value of a Static Random Access Memory (SRAM) cell~\cite{holcomb2009power}, which are uniquely different between devices. 
These properties serve as a unique fingerprint of each device.
Traditional authentication and key exchange using public-key cryptography is not suitable for resource-constrained IoT devices due to the significant computing power required for them. Additionally, they have the requirement of storing the secret key in non-volatile memory, which an adversary may steal, something that the remote and open placement of these IoT devices is susceptible to~\cite{suh2007physical}. 
PUFs, on the other hand, are natural fingerprints for devices, are unique to each device, and are lightweight as well. 

\begin{figure}[b]
\centering
\includegraphics[width=0.8\columnwidth]{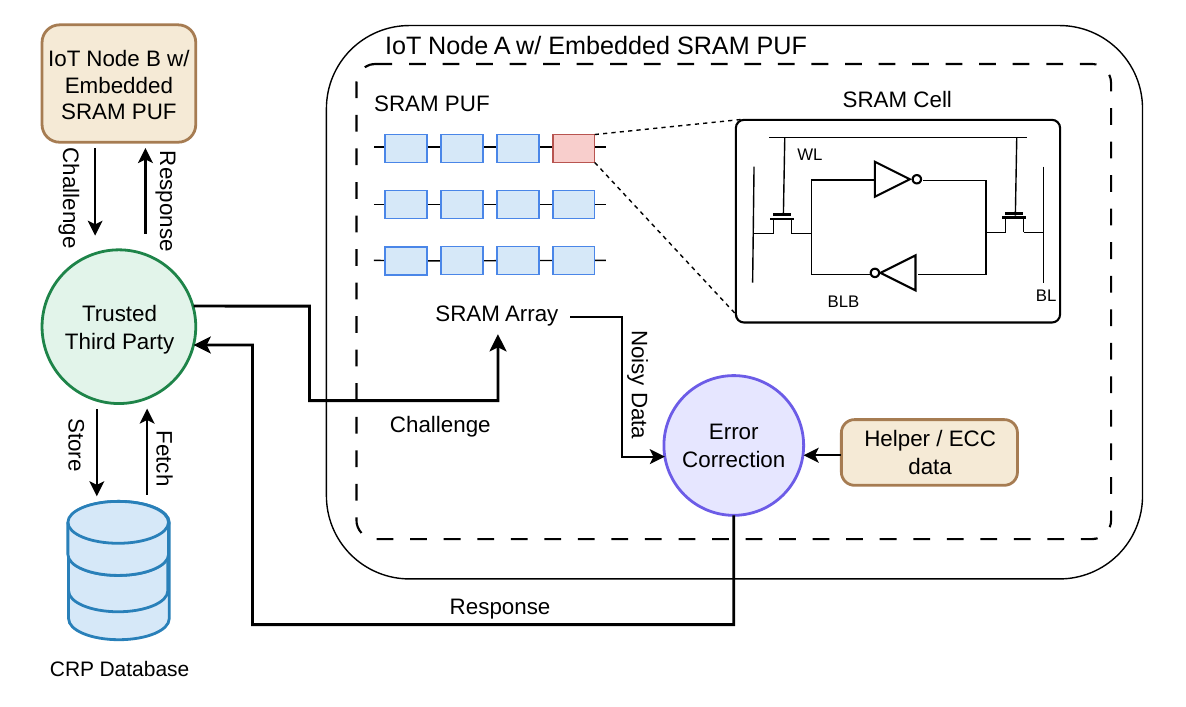}
\caption{PUF-based IoT system utilizing SRAMs.}
\Description{IoT system consisting of a Trusted-Third Parity sending challenges and collecting response from two nodes. Each node has an embedded SRAM array with error-correction.}
\label{fig:sram_iot_system}
\end{figure}

A widely used PUF construction is based on the power-up behavior of SRAM cells. 
As shown in~\autoref{fig:sram_iot_system}, an SRAM cell is formed by two cross-coupled inverters, creating a bistable circuit structure. 
During power-up, random but repeatable manufacturing-induced process variations introduce a small imbalance between the two stable states, causing the cell to initialize preferentially to logic-high (1) or logic-low (0). 
This intrinsic start-up state serves as a unique hardware fingerprint and forms the basis of SRAM PUF operation~\cite{holcomb2009power}.
The SRAM is a 2-D array of such cells, with as many columns as its output data width and as many rows as it has addresses.
Its use as a PUF therefore follows its use as memory: the address maps to the challenge and the data readout maps to the response~\cite{guajardo2007fpga}.

Since the memory size of SRAMs is in the range of kilobytes to megabytes, the number of total challenges, which is equal to the number of addresses, is in the range of $2^{10}$ to $2^{30}$, which is not enough for it to be classified as a strong PUF and thus makes it a weak PUF and not suitable for authentication \cite{ruhrmair2010modeling}.
Although the physical entropy of the SRAM is limited to the size of its memory, and there is no possible way to improve a weak PUF into a strong PUF, this paper designs, implements, and evaluates an enhanced challenge-application and response-extraction mechanism, Device-Unique Feedback Traversal (DUFT), that extracts a response of estimated min-entropy and provides a larger logical input space through the wrapper. 
The main contributions of this paper are summarized as follows:
\begin{itemize}
\item \textbf{A device-unique address traversal mechanism} that derives each round's address from feeding back the device's stability-mask fingerprint output, the device's PUF array hash, the input traversal-seed, and round number for a unique and secret traversal.
\item \textbf{An entropy-estimated response-sizing method} that uses the device's measured per-bit min-entropy and measured error statistics to determine the number of rounds for our feedback-traversal wrapper for a target min-entropy response applied to our two reference hardware.
\item \textbf{A lightweight mutual authentication protocol} based on the entropy-estimated response requiring no pre-shared key or public-key cryptography.
\item \textbf{A PUF-based key-encapsulation protocol} that uses the entropy-estimated response to support secure key establishment on resource-constrained nodes.
\item \textbf{An end-to-end hardware realization} of DUFT and the resulting mutual authentication protocol on physical IoT nodes on a discrete off-chip SRAM platform and an embedded on-chip SRAM microcontroller with performance evaluation.
\end{itemize}

The remainder of this paper is organized as follows.
\autoref{sec:backgnd} introduces background and related work to the paper. 
\autoref{sec:proposed} goes over DUFT and the proposed protocol. 
\autoref{sec:securityanalysis} provides a detailed security analysis of the DUFT mechanism, protocol, and entropy estimation. 
\autoref{sec:Methodology} presents the hardware implementation, experimental setup, and evaluation results, and~\autoref{sec:Conclusion} provides the conclusion.

\section{Background and Related Work}
\label{sec:backgnd}
\subsection{Background}
\subsubsection{Enrollment and Authentication}
A PUF is commonly modeled as a physical challenge-response function, where an applied input is referred to as a challenge and the corresponding output is referred to as a response. 
Although two PUF instances may be fabricated using the same design and manufacturing process, intrinsic process variations cause each instance to generate device-specific responses for the same challenge. 
As a result, the resulting challenge-response pairs serve as unique hardware fingerprints for device identification and authentication~\cite{herder2014physical}. 
PUFs are utilized in two main stages:
\begin{itemize}
    \item Enrollment:
    For the purpose of authentication, the first stage is to collect multiple challenge-response pairs  from a device in a completely secure environment and store them for future use in a database.
    \item Authentication:
    Finally, when it is time to authenticate that a device is who it claims to be, we simply send a challenge from the stored database and compare the device's response with our stored response~\cite{suh2007physical}.
\end{itemize}

\subsubsection{Weak vs. Strong PUFs}
One popular way to characterize physical unclonable functions is by the size of their challenge-response pair (CRP) space. In this characterization, they fall broadly into two categories: strong PUFs and weak PUFs~\cite{ruhrmair2010modeling}.
Delay-based PUFs such as arbiter PUFs have a challenge space that grows exponentially with the number of delay stages, giving them a massive challenge space and thus making them prime examples of strong PUFs.
Memory-based PUFs, contrastingly, due to the large silicon area taken up by memory cells, have a challenge space limited by the total amount of memory they can occupy. This challenge space can be computationally exhausted by an adversary relatively easily, and therefore they are classified as weak PUFs.

\subsubsection{Reliability and Error Correction}
Depending on the design of a PUF and the way they are excited, there is an inherent error rate associated with them. 
Arbiter PUFs generate responses by comparing the signal delay of two wires by making an electrical signal propagate through them. 
SRAM PUFs observe the value that an SRAM cell settles into after powering up. 
Although manufacturing variation creates the bias that gives a PUF its identity, process randomness can lead to errors, and both examples are further influenced by environmental conditions such as temperature.
Effective error correction is therefore vital to a reliable response~\cite{dodis2004fuzzy}.
The standard approach, shown in~\autoref{fig:sram_iot_system}, stores helper data against each response to correct random errors on the fly.

\subsubsection{Fuzzy Extractor}
A fuzzy extractor is a cryptographic primitive used to derive a reliable, secret key from a noisy physical source, such as a PUF. 
As formalized by Dodis et al.~\cite{dodis2004fuzzy}, a fuzzy extractor executes two primary functions: error correction and privacy amplification. While error correction leverages the helper data algorithms discussed in the previous subsection, privacy amplification mitigates the physical imperfections of the silicon. 
In a computational fuzzy extractor, privacy amplification passes the raw data, which carries manufacturing bias and spatial correlation among neighbouring cells, through a cryptographic hash whose avalanche and diffusion properties destroy structural leakage, transforming the biased physical response into a computationally uniform output.

\subsubsection{Controlled PUF and Hash Composition}
The controlled PUF architecture, introduced by Gassend et al.~\cite{Gassend2002Controlled}, combines a cryptographic hash function with a PUF to prevent an adversary from directly observing raw challenge-response pairs by hashing the responses of the PUF. 
This makes modeling attacks harder without requiring changes to the underlying PUF hardware. 
Our DUFT mechanism only makes use of the controlled PUF principle of hashing the final response to prevent the adversary from observing multi-round power-up values of the SRAM.
The seed-based feedback device-unique traversal by utilizing the unique unstable and stable bit positions, and the device array hash as detailed in~\autoref{sec:proposed}, is to the best of the authors' knowledge a novel mechanism, separate from the controlled-PUF literature. 


\subsection{Related Work}
\subsubsection{SRAM PUF Wrappers}
Recent works have attempted to make SRAM PUFs usable for authentication of IoT nodes by implementing wrappers for response-extranction and creating a larger input space to make adversary attacks that exhaust the challenge space infeasible. 
Garrard et al.~\cite{garrard2024xorsram} XOR the outputs of two distinct SRAMs, so the input space becomes the product $C_{SRAM1} \times C_{SRAM2}$ of the two individual spaces.
However, this is not suitable, as it needs more SRAMs to obtain a sufficiently large input space and there is no accounting for the entropy of the response or the error-correction. 
Hu et al.~\cite{hu_spseap} implement SPSE-AP to expand the logical CRP space of an SRAM through permutation of a single startup SRAM reading. 
DUFT, on the other hand, creates a device unique secret address access sequence for every challenge input.
Further, each challenge in DUFT uses a fixed fraction of the array's entropy rather than the complete array, and we also provide a concrete experimentation-driven entropy-estimation, an error-correction scheme and obfuscation through our fuzzy extractor design. 
SPSE-AP does not provide experimentation for error rate.

Baek et al.~\cite{Baek_reconfigurable} present a reconfigurable SRAM PUF that cascades pull-up and pull-down transistors into each cross-coupled inverter, each challenge bit selecting which transistor is applied, reaching $1.6 \times 10^8 \approx 2^{27}$ challenges on a custom ASIC.
Increasing it further would require a larger chip.
DUFT creates a significantly larger input space  wrapper with commercially available SRAMs, parameterized to be increased or decreased along with a entropy-estimate based parameterization for response extraction.

\subsubsection{PUF-based authentication protocols}
Several works in the literature on PUF-based protocols exist \cite{MallPUFsurvey}, and most generally follow the standard of communicating IoT nodes first being authenticated and then exchanging keys with the help of a trusted third party, as shown by~\autoref{fig:sram_iot_system}.
However, most of these works suffer from some significant shortcomings: many assume a pre-shared key to keep the PUF-authentication protocol encrypted and prevent an eavesdropper from obtaining challenges; many rely on a Trusted Third Party (TTP) for a significant part of the protocol; and many rely on computationally expensive cryptographic primitives such as elliptic curve cryptography, causing the lightweight use case of PUFs to lose its purpose.

For example, Farha et al.~\cite{Farha_100} expand the challenge space by enrolling a 100-bit response and sending a random permutation of those 100 locations as the challenge, but rely on a pre-shared network key to hide the challenge, permutation, and response obfuscation, and on a single 100-bit response as their entropy source.
DUFT assumes no pre-shared key, and the feedback ensures every challenge accesses a unique sequence of cells.

Chatterjee et al.~\cite{chatterjeepuf} use PUF responses as public identity within a Weil/Tate pairing scheme over elliptic curves, and Qureshi and Munir~\cite{qureshi2022pufrake} obfuscate CRPs with a Fisher--Yates bit-shuffling mask but use public-key cryptography for key exchange. Both lose the lightweight advantage that motivates PUF-based protocols, which DUFT retains by avoiding PKC entirely.
Nimmy et al.~\cite{nimmy2023puf} address machine-learning attacks by storing challenges as pairs via geometric threshold secret sharing, avoiding a challenge database, but employ no error correction, assume an ideal PUF, and reuse a single response per device.
DUFT supplies a non-repeating stream of logical CRPs with experimentally validated error correction.

Moreover, Delvaux et al.~\cite{Delvaux2015survey} surveyed numerous PUF-based protocols and identified that many designs violate a fundamental design principle of PUF systems: secret information should not be permanently stored on the device. 
These limitations suggest that an effective PUF-based security mechanism should eliminate permanent secret storage, maintain lightweight operation, provide a sufficiently large challenge space, and support mutual authentication. 
While hash-based composition with PUFs has been discussed at a theoretical level~\cite{Delvaux2015survey, Gassend2002Controlled}, no prior work realizes a concrete feedback mechanism over SRAM PUFs with experimentally validated error correction and a full authentication protocol.

These limitations highlight the need for a lightweight SRAM PUF based wrapper desgin that avoids permanent secret storage while providing an expanded logical input space, reliable response generation, and secure authentication and key establishment. 
\autoref{tab:related_comparison} compares features of our proposed design with related works we discussed in this section.

\begin{table}[htbp]
\centering
\caption{Comparison of PUF-based authentication schemes and SRAM-PUF wrappers.
Schemes are identified by citation. \ding{51}~=~satisfied; \ding{55}~=~not
satisfied; --~=~not applicable, not reported, or not directly comparable.}
\label{tab:related_comparison}
\small
\setlength{\tabcolsep}{5pt}
\renewcommand{\arraystretch}{1.15}
\begin{tabular}{lcccccccc}
\toprule
 & \cite{garrard2024xorsram} & \cite{hu_spseap}
 & \cite{Baek_reconfigurable} & \cite{Farha_100}
 & \cite{chatterjeepuf} & \cite{nimmy2023puf}
 & \cite{qureshi2022pufrake} & \textbf{Ours} \\
\midrule
PUF type                & SRAM & SRAM & SRAM & SRAM & Arb. & Gen. & Arb. & \textbf{SRAM} \\
\midrule
DB/CRP exhaustion prot. & \ding{55} & \ding{51} & \ding{51} & \ding{51} & -- & \ding{55} & -- & \textbf{\ding{51}} \\
Response flexibility    & \ding{55} & \ding{51} & \ding{55} & \ding{51} & \ding{55} & \ding{55} & \ding{55} & \textbf{\ding{51}} \\
Error correction        & --        & \ding{55} & \ding{51} & \ding{51} & \ding{51} & \ding{55} & \ding{51} & \textbf{\ding{51}} \\
Mutual authentication   & --        & \ding{51} & --        & \ding{55} & \ding{51} & \ding{51} & \ding{51} & \textbf{\ding{51}} \\
Key establishment       & --        & \ding{51} & --        & \ding{55} & \ding{51} & \ding{51} & \ding{51} & \textbf{\ding{51}} \\
No pre-shared key       & --        & \ding{51} & --        & \ding{55} & \ding{51} & \ding{51} & \ding{51} & \textbf{\ding{51}} \\
No PKC                  & --        & \ding{51} & \ding{51} & \ding{51} & \ding{55} & \ding{51} & \ding{55} & \textbf{\ding{51}} \\
\midrule
Security (bits)         & --        & --        & 80+       & 128       & 128 & -- & 64 & $\mathbf{128^{\dagger}}$ \\
\bottomrule
\end{tabular}
\smallskip
\begin{minipage}{\columnwidth}
\footnotesize
PKC: public-key cryptography. DB/CRP exhaustion prot.: database/CRP exhaustion
protection. Response flexibility: flexibility of response size and its
min-entropy. Arb.: arbiter; Gen.: generic.\par
$^{\dagger}$DUFT's security level is a computed function of measured source
entropy and measured worst-corner error rate~\eqref{eq:nr-conditional};
128 is the \emph{target} met by our reference hardware under the assumptions
of~\autoref{subsec:respsecurity}, not a hardware-independent guarantee.
\end{minipage}
\end{table}

\section{The Proposed Mechanism and Protocol}
\label{sec:proposed}
This section presents the proposed DUFT mechanism and its associated authentication and key-establishment protocols.

\subsection{Response Extraction and Enlarging the Wrapper Input Space}
\begin{figure}[t]
\centering
\includegraphics[width=\columnwidth]{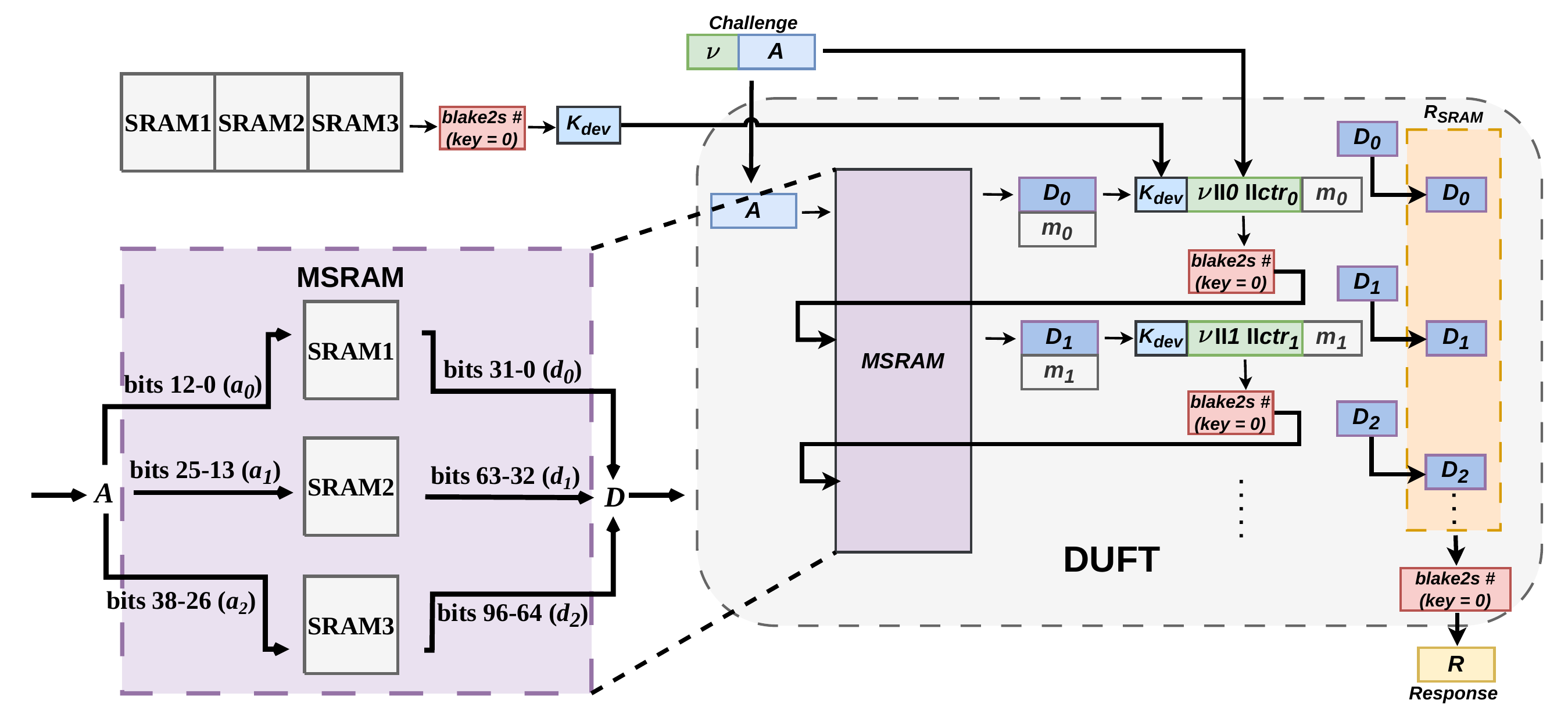}
\caption{DUFT mechanism for iterative challenge generation and response accumulation.}
\Description{DUFT mechanism overview that shows MSRAM functionality, $K_{dev}$ derivation, and the round-by-round feedback and response accumulation.}
\label{fig:rtcr-msram}
\end{figure}

In this section, we outline a wrapper around commercially available SRAMs, or on-board embedded SRAMs, that provides an increased logical input space of the PUF-wrapper and parameterized flexibility in response extraction to to target a estimated amount of min-entropy.
First, a suitable number of SRAMs are chosen.
We use three here, each with a data width of 32 bits and an address width of 13 bits, but DUFT is flexible enough to use more or fewer of either.
Their address widths are combined, and their output data widths as well.
A single SRAM is modeled as the following function:
    \begin{equation}
        d = \mathrm{SRAM}(a)
        \label{eq:SRAM_function}
    \end{equation}
where $d$ is the power-up data of a single SRAM and $a$ is the address provided to it.
Since there are three SRAMs, the combined address $A$ consists of the addresses $a_x$ for all three concatenated together:
    \begin{equation}
        A = a_{3} \parallel a_2 \parallel a_1
        \label{eq:address_concatenation}
    \end{equation}
The total address width of the three SRAMs is therefore $13+13+13 = 39$ bits.
Similarly, the power-up data $D$ from multiple SRAMs is obtained by concatenating their individual power-up data $d_x$, as~\autoref{fig:rtcr-msram} shows:
    \begin{equation}
        D = d_3 \parallel d_2 \parallel d_1
        \label{eq:powerup_data_concatenation}
    \end{equation}
Together the data width of three SRAMs is $32+32+32=96$ bits in this case. 
This combined address being fed to multiple SRAMs to obtain a combined data output is modeled with the following function:
    \begin{equation}
        D = \mathrm{MSRAM}(A)
        \label{eq:MSRAM_equation}
    \end{equation}
    where
    \begin{equation}
        \mathrm{MSRAM}(A) = \ldots \parallel \mathrm{SRAM}(a_2) \parallel \mathrm{SRAM}(a_1) 
        \label{eq:MSRAM_equation_expansion}
    \end{equation}
We introduce some variables now that will be used in the next step. 
Alongside the power-up data, every MSRAM address also has a 96-bit stability mask, $m$, associated with it, recording for each bit position whether the SRAM cell there powers up consistently or unpredictably. 
Its derivation is described in~\autoref{subsec:errcorrenroll}.
For this section it is enough to know that $m$ is a fixed property of an address on a particular device, may be stored on device or transmitted from the server database after enrollment, and discloses only which cells are stable, not what values they hold. 
Which cells are stable and which are unstable is itself a manufacturing artifact: the same process variation that decides whether a cell powers up to 0 or to 1 also decides how strongly it does so, and the weakly biased cells are the ones that unstable in the stability mask.
Two chips of the same part number therefore have different stable-cell patterns, just as they have different power-up values, so the mask is a device fingerprint in its own right, separate from the response and reveals nothing about it.
Next, we introduce the hash of the combined SRAM arrays after error-correction to create a device key $K_{dev}$.
This is a unique value with any instance of an MSRAM setup since every MSRAM will have unique SRAM arrays due to their PUF qualities:
    \begin{equation}
        K_{dev} = \mathrm{BLAKE2s}\big(k = 0,\; d = \mathrm{SRAM_3} ||\mathrm{SRAM_2} || \mathrm{SRAM_1})
        \label{eq:devkey}
    \end{equation}
where $\mathrm{SRAM_{x}}$ are the complete SRAM arrays.
$K_{dev}$ is a function of the array's values and not of any challenge, so it is the same at every power cycle and the same at enrollment as at authentication.
It lives in volatile memory for the power cycle and is never transmitted or written to non-volatile memory.
It is not an entropy source and we do not treat it as one.
Its only purpose is to put the address schedule out of reach of an adversary who holds the public variables involved in the feedback but not the array.
Thirdly, we introduce a 64-bit traversal seed $\nu$ that we will utilize in our feedback traversal.

The core novelty of this paper is now introduced. The device key, the traversal seed, and the round number are concatenated with the round's stability mask and hashed using the unkeyed BLAKE2s-128 function~\cite{aumasson2013blake2}, and the lower $w_A$ bits of the digest are taken as the next MSRAM address:
    \begin{equation}
        h_{r} = \mathrm{BLAKE2s}(k = 0, d = K_{dev} \parallel \nu \parallel r \parallel m_{r})
        \label{eq:roundhash}
    \end{equation}
    \begin{equation}
        A_{r+1} = h_{r}\mod 2^{w_A}
        \label{eq:round_addr}
    \end{equation}
Here $w_A$ is the width of the MSRAM address and $r$ is the round number, iterating from $0-(n_r-1)$ where $n_r$ is the number of rounds.
The steps are now repeated for the next round.
The next round's MSRAM data is read out:
    \begin{equation}
        D_{r} = \mathrm{MSRAM}(A_{r})
        \label{eq:round_data}
    \end{equation}
Feeding back the round data $D_r$ would be the more obvious choice, and we do not do it.
\autoref{subsec:feedbackdesign} gives a detailed explanation for the choice of $K_{dev}$ and $m_r$ to to drive the feedback.

We make another addition to the hash-calculation at this point.
To ensure that no round within a single challenge's traversal revisits a per-chip sub-address already sampled earlier in the same traversal, a collision-check step is introduced before a candidate address $A_r$ is accepted. 
Each candidate address is split into its per-chip sub-addresses $(a_1, a_2, a_3)$ as in~\eqref{eq:address_concatenation}, and checked against the sets of sub-addresses already visited by each SRAM in the current session, $v_1$, $v_2$, $v_3$. 
If any sub-address $a_x$ has already been visited, the candidate is rejected and a fresh address is re-derived by incrementing a per-round counter $ctr_r$ and recomputing the hash:
    \begin{equation}
        h_{r} = \mathrm{BLAKE2s}(k = 0, d = K_{dev} \parallel \nu \parallel r \parallel ctr_{r} \parallel m_{r})
        \label{eq:roundhash_counter}
    \end{equation}
This is repeated, incrementing $ctr_r$ each time, until a candidate address is found whose per-chip sub-addresses are all unvisited within the current session, at which point $A_{r+1}$ is accepted and each $v_i$ is updated with the newly accepted $a_i$.
This ensures that within-session per-chip sub-address collisions cannot occur for any traversal length $n_r$ given that $n_r$ is smaller than the address-space of a single SRAM.

In each round, MSRAM outputs are accumulated and concatenated to form a larger response, as demonstrated by~\autoref{fig:rtcr-msram}.
    \begin{equation}
        R_{SRAM} =  \ldots \parallel D_2 \parallel D_1 \parallel D_0
        \label{eq:sramresponse}
    \end{equation}
This process is repeated for $n_r$ rounds until the accumulated response carries sufficient min-entropy.
$n_r$ is not a fixed design parameter but is computed from the empirically estimated per-bit min-entropy and error-rate of the deployed source, as derived in~\eqref{eq:nr-conditional}.
Although each round contributes 96 raw bits, only stable positions carry entropy and a further deduction to entropy is owed to the published error-correction data, so $n_r$ is considerably larger than just dividing the target by 96.

Finally, the accumulated response is hashed using the  BLAKE2s-128 unkeyed function~\cite{aumasson2013blake2} to obfuscate the SRAM inputs and serve as the fuzzy extractor privacy amplification step:
    \begin{equation}
        R = \mathrm{BLAKE2s}(k=0, d=R_{SRAM})
        \label{eq:blakehash}
    \end{equation}
The PUF design is then formalized with these equations:
    \begin{equation}
        C = \nu \parallel A
        \label{eq:challenge}
    \end{equation}
    \begin{equation}
        R = \mathrm{PUF}(C)
        \label{eq:puf_equation}
    \end{equation}
The use of BLAKE2s in every round as an intermediate hash $h_r$ creates a randomized and deterministic chain from the round mask $m_r$ to the next address $A_{r+1}$, unique to the traversal seed $\nu$, the starting address $A$, the stable-cell pattern of the chips, and the device key $K_{dev}$.
Three properties of the SRAM therefore do separate work: the positions of the stable cells select the path, the values of the array as a whole key that selection through $K_{dev}$, and the values of the selected cells supply the entropy of the response.
Modeled as a random oracle, BLAKE2s has no exploitable correlation between its input and output, so distinct seeds give decorrelated paths over the same array.

The final hash, $R_{SRAM} \to R$ serves a different purpose. 
It obfuscates the accumulated multi-round power-up data, similar to how it is done with the single response in controlled-PUF literature~\cite{Gassend2002Controlled}, while simultaneously acting as the privacy-amplification step of the fuzzy extractor, transforming the biased physical response into a computationally uniform key. We also clarify here that all three BLAKE2s hashes used in DUFT for generating the device key, for computing the round address, and the final response hash are the unkeyed BLAKE2s hash-variant with a 128-bit digest.

\subsection{Error Correction and Enrollment}
\label{subsec:errcorrenroll}
Error correction with SRAM PUFs is a challenging problem. 
Firstly, there are normal random errors that can occur arbitrarily with random probability. 
Secondly, with SRAMs, cells are classified into two categories with regards to errors~\cite{holcomb2009power}: stable cells, which are SRAM cells or bits that almost always power up to 0 or 1, and unstable cells, which are unpredictable in what they power up as. 
For this use case, unstable cells in SRAMs have no utility due to a lack of reliability, which is a requirement of a PUF.
Therefore, we utilize the position of these cells, which is a unique fingerprint per-chip for the feedback traversal.
Stable cells are of interest and will be the subject of our error correction to treat random errors.

The design follows a two-stage approach: first, unstable bits of the response will be eliminated using a masking scheme~\cite{maes2009low}.
After that, random errors in the stable bits will be corrected using Bose–Chaudhuri–Hocquenghem (BCH) codes~\cite{bose1960class}.
Korenda et al.~\cite{korenda2019proof} implement various variants of this two-stage approach for SRAM PUFs while masking of unstable bits was a strategy first introduced by Holcomb et al.~\cite{holcomb2009power}.
Referring to the bit-width a round MSRAM read, $D_r$, as $w_D$, masking this $w_D$-bit read removes $l_m$ unpredictable positions of unstable bits and leaves $s = w_D - l_m$ stable bits, and only those $s$ bits carry any entropy or need correcting.

Both the stability mask and the error-correction parity are properties of the individual SRAM arrays rather than of any one challenge, so we generate both once per device and keep both in device non-volatile memory.
For our configuration the mask is $3 \times 2^{13} \times 32$ bits, or 96\,KB.
The mask records which positions are reliable and not what values they hold, and the parity leaks at most its own length in min-entropy, the leakage in min-entropy by both as $l_m$ and $l_{BCH}$ are described in~\autoref{subsubsec:entropy-definitions}.

We generate parity bits across the complete array and not per-challenge due to necessity and not as a choice.
BCH parity is linear in the data it protects, so each parity bit is a known combination of specific golden-reference bits.
Even though our traversal is secret, an adversary who knows the address schedule of every challenge also knows which array locations physical bits went into each combination.
Publishing parity per challenge would hand him one system of linear equations per CRP over the same unknown array.
For our reference hardware (\autoref{sec:Methodology}), our array has stable bits in the range of a megabit and a per-traversal sketch would supply 100s of equations each, so a few thousand enrolled CRPs would be enough to solve for the whole array without any physical access.
Generating the parity once per device removes the per-challenge equations entirely, and what remains is a single fixed leak against the array as a whole.

We correct the array in windows rather than in one block.
We choose a window to be $win = 512$ consecutive addresses on one chip, so $512\times32 = 16{,}384$ raw bits, of which $s/w_D \times100\%$ are stable as an estimate.
The choice of $win$ is a trade-off.
A single code over the whole array would need to correct errors in the thousands, and BCH decoding costs $O(n \cdot t)$ in the Chien search for a code correcting $t$ errors for a codeword of size $n$, which could be too much for a microcontroller.
Very small windows decode quickly but spend a large fraction of each block on parity, and since a stable bit carries well under one bit of min-entropy, that fraction comes straight out of the entropy budget.
Five hundred and twelve addresses is a suitable middleground.

To choose the BCH variant for a window we need $t$, the number of flips it must absorb.
Sizing $t$ from a per-bit error probability assumes flips fall independently, so that the error count over a window is Poisson and its spread is the square root of its mean.
Our measurements in~\autoref{subsubsec:error-corr-experiments} show this does not hold at window scale: flips cluster across addresses, with a variance about twice the mean on the Cypress and about five times on the STM32F407, our two reference hardware.
We therefore size $t$ from the measured mean $\mu$ and standard deviation $\sigma$ of the window error count directly, rather than from a per-bit rate and an independence assumption:
\begin{equation}
    t = \lceil \mu + c\sigma \rceil .
    \label{eq:t-values}
\end{equation}
Here $c$ has to cover all $48$ windows jointly and not one of them, since a single window failing to decode spoils the whole restore.
We take $c = 6$, and we evaluate~\eqref{eq:t-values} at the upper end of the $95\%$ confidence intervals on $\mu$ and $\sigma$ rather than at the point estimates, so that the sizing survives the sampling error of the measurement itself.
This gives $t = 23$ for the Cypress and $t = 53$ for the STM32F407.
The two differ because the parts differ: the STM32F407 has both a higher mean error count and a wider spread, so it needs more than twice the correction capability for the same window size.

A binary BCH code over GF($2^{w_f}$) spends $w_f$ parity bits per correctable error and has a block length of $2^{w_f}-1$, so $w_f$ is the smallest degree whose block length holds the window and its parity together.
Fourteen is too small, since $2^{14}-1$ is short of the $16{,}384$ information bits on their own.
We therefore use shortened variants over GF($2^{15}$): BCH$(16729, 16384, 23)$ for the Cypress, spending $345$ parity bits per window, and BCH$(17179, 16384, 53)$ for the STM32F407, spending $795$.
With $16$ windows on each of the three chips this is $l_{BCH} = 16{,}560$ bits or $2.0$\,KB for the Cypress and $l_{BCH} = 38{,}160$ bits or $4.7$\,KB for the STM32F407.

At enrollment the device reads the array $n_{enroll}$ times, marks the stable positions in $M_{tab}$, records their values in $G_{tab}$, encodes $G_{tab}$ window by window, and writes $M_{tab}$ and the parity table $P_{tab}$ to flash.
$G_{tab}$ is the raw fingerprint of the device and is discarded once the parity is written.
At every power-up the device reads the array once, zeros the unstable positions with $M_{tab}$, decodes each window against $P_{tab}$, and writes the corrected values back over the array in place.
The restored copy costs no extra memory, since it replaces power-up data the device has already consumed, and it lives only in volatile memory for that power cycle.
Traversals then read the restored array directly, so no correction happens on the authentication path at all.

It must be stated that the parameters above are derived from measurements of our reference hardware and are not standard numbers to be plugged in.
DUFT implementations must evaluate their underlying physical source before choosing a window size and a BCH variant.
\autoref{sec:securityanalysis} makes this more clear.

\begin{algorithm}[!htbp]
\scriptsize
\caption{DUFT mask-table, parity-table and Helper Functions.}
\label{alg:enrollcrp}
\KwIn{$n_{SRAM}, n_{enroll}, n_r$: number of SRAMs, enrollment power cycles, rounds}
\KwIn{$win, w_f, t$: window size in addresses, field degree, and errors corrected per window}
\KwIn{$ID$: 20-bit unique Node ID; $C$: complete challenge for the PUF}
\KwOut{$M_{tab}, P_{tab}$: stability mask and window parity tables, kept in device NVM}
\KwOut{$CRP$: challenge-response pair as stored in the server database}
\KwData{$arr_j[a]$: per-bit running sum for chip $j$ at sub-address $a$}
\KwData{$G_{tab}, G$: array and traversal golden reference, session memory only}
\KwData{$W$: one window of masked array data during restore}
\KwData{$Z$: Complete restore $n_{SRAM}$ array concatenated in series}
\KwData{$a_j$: chip $j$'s sub-address; $v_j$: chip $j$'s visited sub-addresses}
\KwData{$ctr_r$: per-round address-retry counter}
\KwData{$K_{dev}$: device key, volatile, recomputed each power cycle}
\BlankLine

\Fn{\FEnrollTables{$n_{enroll}$}}{
  $arr_j[a] \gets [0,0,\ldots] \;\; \forall j \in [1, n_{SRAM}], \; \forall a \in [0, 2^{w_a}{-}1]$\label{line:initsum}\;
  \For{$i = 1$ \KwTo $n_{enroll}$}{
    \FPowerCycle{}\label{line:powercycle}\;
    \For{$a = 0$ \KwTo $2^{w_a}-1$}{
      $d_{n_{SRAM}} \parallel \cdots \parallel d_1 \gets{}$\FMSRAM{$a \parallel \cdots \parallel a$}\label{line:readall}\tcp*{one sub-address on every chip}
      $arr_j[a] \gets arr_j[a] + {}$\FtoBitArray{$d_j$} \textbf{ for } $j = 1$ \textbf{ to } $n_{SRAM}$\label{line:accumecc}\;
    }
  }
  \For{$j = 1$ \KwTo $n_{SRAM}$, $a = 0$ \KwTo $2^{w_a}-1$}{
    $M_{tab}[j][a] \gets (arr_j[a] = n_{enroll}) \lor (arr_j[a] = 0)$\label{line:maskecc}\;
    $G_{tab}[j][a] \gets (arr_j[a] = n_{enroll})$\label{line:golden}\;
  }
  \For{$j = 1$ \KwTo $n_{SRAM}$, $k = 0$ \KwTo $2^{w_a}/win-1$}{
    $P_{tab}[j][k] \gets{}$\FWindowEncode{$G_{tab}[j][\,k  win : (k{+}1)  win\,]$}\label{line:winenc}\;
  }
  \textbf{persist} $M_{tab}, P_{tab}$ to NVM; \textbf{zeroize} $G_{tab}$\label{line:persist}\;
  \KwRet $M_{tab}, P_{tab}$\;
}
\BlankLine

\Fn{\FWindowEncode{$G_{win}$}}{
  $c \gets G_{win} \cdot \mathbf{\Gamma}_{BCH}$\label{line:encode}\;
  \KwRet $c[\,winw_d : winw_d + w_f t\,]$\label{line:parity}\tcp*{parity only; the information block is discarded}
}
\BlankLine

\Fn(\tcp*[f]{once per power cycle, before anything else}){\FRestore{}}{
  \For{$j = 1$ \KwTo $n_{SRAM}$, $k = 0$ \KwTo $2^{w_a}/win-1$}{
    $W \gets 0$\;
    \For{$a = kwin$ \KwTo $(k+1)win -1$}{
        $W \gets W \parallel (SRAM_{j}(a) \And M_{tab}[j][a])$\label{line:winread}\;
    }
    $W \gets{}$\FBCHCorrect{$W, P_{tab}[j][k]$}\label{line:windec}\;
    \textbf{write} $W$ back over addresses $k win$ to $(k{+}1) win-1$ of chip $j$\label{line:winwrite}\;
  }
  $Z \gets{}$\FRestored{$SRAM_{n_{SRAM}}$}$ \parallel \cdots \parallel{}$\FRestored{$SRAM_{0}$}\label{line:devkeyconcat}\tcp*{Read out complete restored arrays}
  $K_{dev} \gets{}$\FBLAKE{$k = 0, d = Z$}\label{line:devkey}\;
  \KwRet $K_{dev}$\;
}
\BlankLine

\Fn(\tcp*[f]{read the corrected array, shaped like \textsc{MSRAM}}){\FRestored{$A$}}{
  $a_{n_{SRAM}} \parallel \cdots \parallel a_1 \gets A$\label{line:restfn}\;
  \KwRet \FRead{$n_{SRAM}, a_{n_{SRAM}}, 1$}$ \parallel \cdots \parallel{}$\FRead{$1, a_1, 1$}\;
}
\BlankLine

\Fn(\tcp*[f]{mask lookup, shaped like \textsc{MSRAM}}){\FMask{$A$}}{
  $a_{n_{SRAM}} \parallel \cdots \parallel a_1 \gets A$\label{line:maskfn}\;
  \KwRet $M_{tab}[n_{SRAM}][a_{n_{SRAM}}] \parallel \cdots \parallel M_{tab}[1][a_1]$\;
}
\end{algorithm}
 
\textbf{Pseudocode:}
Algorithm~\ref{alg:enrollcrp} depicts mask-table and parity-table enrollment and array restoration.
It comprises five functions: \textsc{Enroll\_Tables}, \textsc{Window\_Encode}, \textsc{Restore}, \textsc{Restored} and \textsc{Mask}.
Throughout, $w_x$ denotes the width in bits of $x$, and quantities spanning several chips are formed by concatenation in chip order, as in~\eqref{eq:address_concatenation} and~\eqref{eq:powerup_data_concatenation}.
 
\textsc{Enroll\_Tables} runs once per device and characterizes the whole addressable array.
It initializes a per-chip, per-address accumulator (line~\ref{line:initsum}), power-cycles the SRAMs $n_{enroll}$ times (line~\ref{line:powercycle}), and on each cycle walks every sub-address, issuing the combined address $a \parallel \cdots \parallel a$ so that one \textsc{MSRAM} read returns sub-address $a$ on all $n_{SRAM}$ chips at once (line~\ref{line:readall}) and adds the result into the running sums (line~\ref{line:accumecc}).
A bit is marked stable in $M_{tab}$ if it powered up consistently to 1 or to 0 across all cycles (line~\ref{line:maskecc}), while the golden reference $G_{tab}$ records only the stable-1 positions (line~\ref{line:golden}).
Parity for each 512-address window is then enrolled into $P_{tab}$ (line~\ref{line:winenc}).
\textsc{Enroll\_Tables} reads the array directly through \textsc{MSRAM} rather than \textsc{Restored}, since it is the step that produces the tables \textsc{Restore} needs.
$M_{tab}$ and $P_{tab}$ are written to non-volatile memory and are assumed public. $G_{tab}$ is the raw fingerprint of the device and is held in volatile memory for the enrollment session only (line~\ref{line:persist}).

\textsc{Window\_Encode} takes one window of the golden reference, encodes it with the generator matrix $\mathbf{\Gamma}_{BCH}$ (line~\ref{line:encode}) and returns only the $w_f \cdot t$ parity bits (line~\ref{line:parity}). The information block is discarded, so each window leaks exactly $w_f \cdot t$ bits and the array as a whole leaks $l_{BCH}$, the sum over all windows~\cite{dodis2004fuzzy}.

\textsc{Restore} runs once per power cycle, before either enrollment or authentication.
For each window it masks the power-up data with $M_{tab}$ (line~\ref{line:winread}), decodes it against $P_{tab}$ (line~\ref{line:windec}), and writes the corrected values back in place (line~\ref{line:winwrite}), costing no extra memory since it overwrites data already consumed.
Once every window is corrected it hashes the restored array into $K_{dev}$ (line~\ref{line:devkey}) and returns it. 
The hash must be performed after the all widnows have been error-corrected, since a key taken over a partly corrected array would differ between power cycles.
Both tables come from non-volatile memory, so nothing has to be transmitted for this to run, and $K_{dev}$ never leaves volatile memory.

\textsc{Restored} is a plain read of the corrected array, split and concatenated the same way \textsc{MSRAM} splits and concatenates the three SRAMs (line~\ref{line:restfn}). The difference is that \textsc{Restored} has already masked and corrected the arrays so the read is error-corrected stable bits PUF data.
This way masking and decoding only has to be done once.
 
\textsc{Mask} is a lookup shaped like \textsc{MSRAM}: it splits a combined address into per-chip sub-addresses, indexes the corresponding table row for each chip, and concatenates the results in chip order (line~\ref{line:maskfn}), so a round mask $m$ has the same $w_D$-bit layout as a round of data $D$ and the two align bit for bit.
 
\subsection{Response Extraction for Enrollement and Authentication}
 
After enrollment mask and parity tables has been performed, a trusted-third-party (TTP) may enroll CRPs from the DUFT wrapper after issuing challenges to the device.
This has to be performed in a secure environment.
After CRP enrollment, the TTP may issue a challenge and the device will generate the response against the challenge.
We only show CRP enrollment and response-generation
Algorithm~\ref{alg:rtcr} via DUFT here.
A complete protocol using this wrapper will be demonstrated later. 
 
\textbf{Pseudocode}:

Algorithm~\ref{alg:rtcr} depicts how the DUFT mechanism extracts the response for enrollment and authentication of a given challenge.
It uses \textsc{Mask} and \textsc{Restored} from Algorithm~\ref{alg:enrollcrp}, defines \textsc{NextAddr} and \textsc{TryVisit} itself, and assumes $n_r$, $t$ and $w_f$ are solved once for the deployed source rather than fixed by the design.

\textsc{Enroll\_CRP} parses $C = \nu \parallel A$, empties $G$ and $v$ (line~\ref{line:parsec}), then iterates for $n_r$ rounds.
Each round prepends the corrected data at the current address to $G$ (line~\ref{line:goldlookup}), so that round $0$ ends up in the low-order position, matching~\eqref{eq:sramresponse}.
The mask at that address is looked up separately (line~\ref{line:masklookup}), since it drives the feedback rather than the response.
No masking or correction happens in this loop, as \textsc{Restore} did both before it ran.
The retry counter is cleared (line~\ref{line:initctr}) and the next address derived from the device key and the round mask by \textsc{NextAddr} (line~\ref{line:nexthash}), which hashes $K_{dev}$, $\nu$, the round number, $ctr_r$ and $m$ together and reduces modulo $2^{w_A}$.
Each candidate is offered to \textsc{TryVisit} (line~\ref{line:retryloop}), which accepts it only if every per-chip sub-address is unvisited this session; if rejected, $ctr_r$ is incremented (line~\ref{line:incrctr}) and a fresh candidate derived, until one is accepted.
No round within an enrollment session can therefore revisit a sub-address used by an earlier round.
After all rounds the response $R$ is produced by hashing $G$ (line~\ref{line:finalr}) and $G$ is zeroised (line~\ref{line:zeroise}).
The entry $[ID, C, R]$ is returned for storage in the server database (line~\ref{line:crpassem}).
 
\textsc{DUFT} walks the identical loop (lines~\ref{line:parse}--\ref{line:final}), accumulating into $R_{SRAM}$ rather than $G$.
Because $m$ is read from the same stored table, and $K_{dev}$ is taken over the corrected array rather than a live one, the address sequence is identical to the one walked at enrollment, so $R_{SRAM}$ here equals $G$ there for the same challenge, and hashing both gives the same $R$ (line~\ref{line:final}).
\begin{algorithm}[!htbp]
\footnotesize
\caption{DUFT response extraction.}
\label{alg:rtcr}
\KwIn{$n_{SRAM}$: number of SRAMs}
\KwIn{$M_{tab}, P_{tab}$: stability mask and window parity tables held in device NVM}
\KwIn{$n_r$: rounds, solved once for the deployed source from \eqref{eq:nr-conditional}}
\KwIn{$K_{dev}$: device schedule key returned by \textsc{Restore}, volatile}
\KwIn{$C$: complete challenge for the PUF}
\KwOut{$R$: the final hashed response from the PUF}
\KwData{$A, m$: combined address and mask for one round}
\KwData{$\nu$: traversal seed; $ctr_r$: per-round address-retry counter}
\KwData{$v$: per-chip visited-address lists, $v = [v_1, \ldots, v_{n_{SRAM}}]$}
\KwData{$R_{SRAM}$: accumulated response from all rounds}
\BlankLine

\Fn{\FNextAddr{$K_{dev}, \nu, r, ctr, m$}}{
  $h \gets{}$\FBLAKE{$k = 0, d = K_{dev} \parallel \nu \parallel r \parallel ctr \parallel m$}\label{line:nextaddr}
  \KwRet $h \bmod 2^{w_A}$\;
}
\BlankLine

\Fn{\FTryVisit{$A, v$}}{
  $a_{n_{SRAM}} \parallel \cdots \parallel a_1 \gets A$\label{line:tvaj}\;
  \For{$j = 1$ \KwTo $n_{SRAM}$}{
    \If{$a_j \in v_j$\label{line:tveq}}{
      \KwRet \textbf{false}, $v$\;
    }
  }
  append $a_j$ to $v_j$ \textbf{ for } $j = 1$ \textbf{ to } $n_{SRAM}$\label{line:tvappend}\;
  \KwRet \textbf{true}, $v$\label{line:tvtrue}\;
}
\BlankLine

\Fn{\FEnrollCRP{$C$}}{
  $\nu \parallel A \gets C$; \quad $G \gets 0$; \quad $v \gets [\,\emptyset, \ldots, \emptyset\,]$\label{line:parsec}\;
  \For{$i = 0$ \KwTo $n_r-1$}{
    $G \gets{}$\FRestored{$A$}$ \parallel G$\label{line:goldlookup}\;
    $m \gets{}$\FMask{$A$}\label{line:masklookup}\;
    $ctr_r \gets 0$\label{line:initctr}\;
    \Repeat{$ok$}{
      $A \gets{}$\FNextAddr{$K_{dev}, \nu, i, ctr_r, m$}; \quad $ctr_r \gets ctr_r + 1$\label{line:nexthash}\label{line:incrctr}\;
      $ok, v \gets{}$\FTryVisit{$A, v$}\label{line:retryloop}\;
    }
  }
  $R \gets{}$\FBLAKE{$k=0, d = G$}; \quad \textbf{zeroize} $G$\label{line:finalr}\label{line:zeroise}\;
  \KwRet $CRP \gets [ID, C, R]$\label{line:crpassem}\;
}
\BlankLine

\Fn{\FDUFT{$C$}}{
  $\nu \parallel A \gets C$; \quad $R_{SRAM} \gets 0$; \quad $v \gets [\,\emptyset, \ldots, \emptyset\,]$\label{line:parse}\label{line:initv2}\;
  \For{$i = 0$ \KwTo $n_r-1$}{
    $R_{SRAM} \gets{}$\FRestored{$A$}$ \parallel R_{SRAM}$\label{line:accum}\;
    $m \gets{}$\FMask{$A$}\label{line:mslice}\;
    $ctr_r \gets 0$\label{line:initctr2}\;
    \Repeat{$ok$\label{line:retryloop2}}{
      $A \gets{}$\FNextAddr{$K_{dev}, \nu, i, ctr_r, m$}; \quad $ctr_r \gets ctr_r + 1$\label{line:feedback}\label{line:incrctr2}\;
      $ok, v \gets{}$\FTryVisit{$A, v$}\label{line:updatev2}\;
    }
  }
  $R \gets{}$\FBLAKE{$k = 0, d = R_{SRAM}$}\label{line:final}\;
  \KwRet $R$\;
}
\end{algorithm}

\subsection{PUF-based protocol utilizing the Entropy-Budgeted Response}
With the implementation of DUFT, we next propose a simple, yet extremely effective, mutual authentication and key-encapsulation protocol in the space for IoT and resource-constrained devices. 
We first create our proposed enrollment scenario consisting of the traditional trusted third party and a pair of IoT nodes that wish to communicate securely with each other as shown in~\autoref{fig:sram_iot_system}~\cite{pappu2002physical,gassend2002silicon}. 
As discussed, a database CRP entry will consist of $(ID_x, C, R)$, where $ID_x$ is the 20-bit ID of the node x, $C$ is the challenge consisting of the first address $A$ and traversal seed $\nu$, and $R$ is the response.
No helper data is stored with the entry or sent with the challenge, since both the mask and the parity tables are held on the device.

\begin{figure}[t]
\centering
\includegraphics[width=\textwidth]{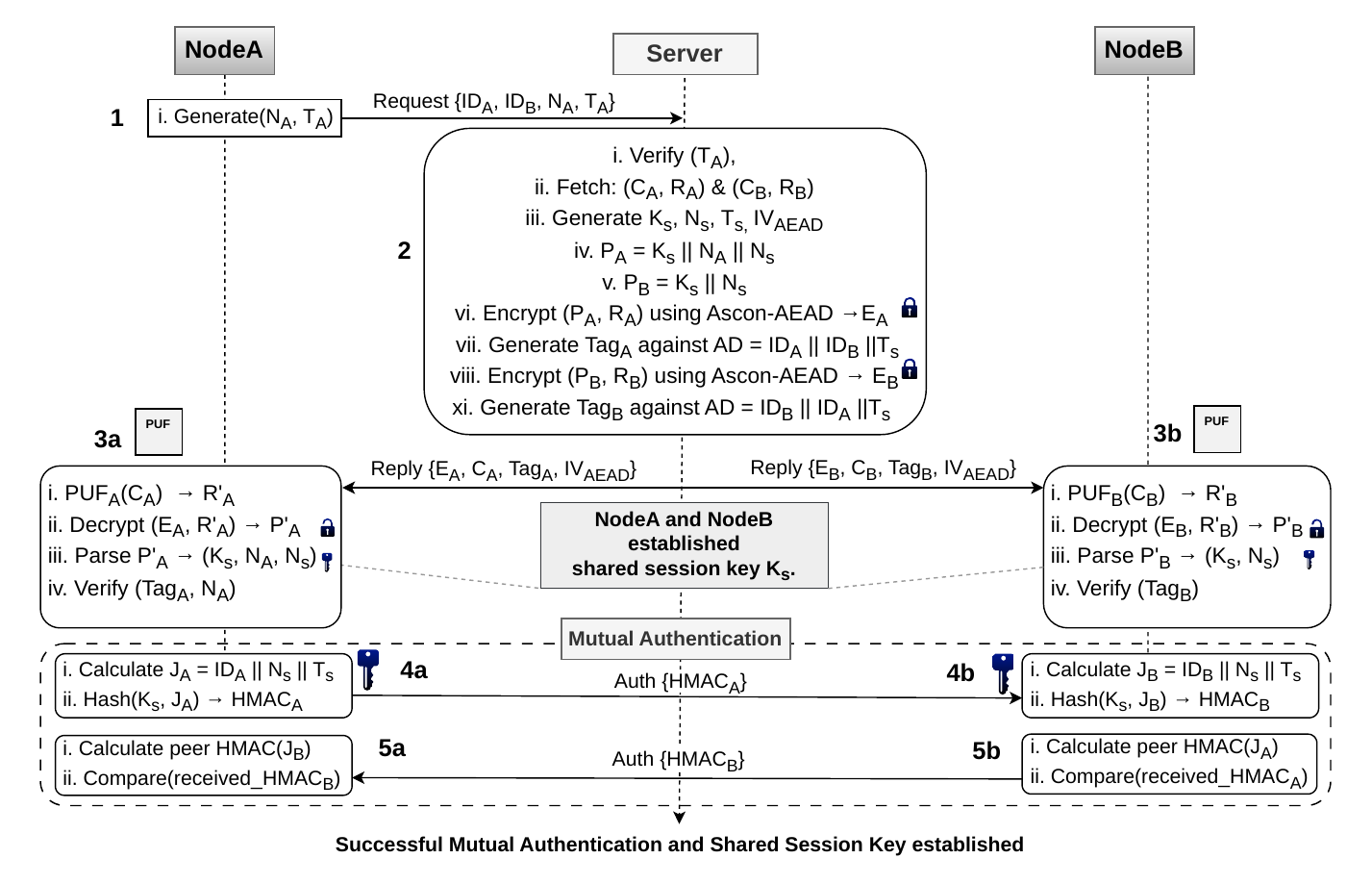}
\caption{Key-exchange and mutual authentication protocol based on our DUFT-PUF.}
\Description{Protocol diagram of the mutual authentication and key-exchange protocol. Three columns to represent the TTP and two nodes. Each column shows the steps of the protocol in sequence.}
\label{fig:protocol}
\end{figure}

\textbf{Steps of the protocol}
\begin{itemize}

\item \textbf{Request (Node A $\to$ Server):} NodeA sends its own NodeID ($ID_A$), the peer node's NodeID ($ID_B$), a freshly generated nonce ($N_A$), and a current timestamp ($T_A$) to the server.

\item \textbf{Server preparation:} The server verifies $T_A$ to prevent replay attacks, then fetches a random CRP entry for each node: $(ID_A, C_A, R_A)$ for NodeA and $(ID_B, C_B, R_B)$ for NodeB. 
It generates a 128-bit session key $K_s$, a session nonce $N_S$, a server timestamp $T_S$, and a unique Authenticated-encryption with Associated-data (AEAD) initialization vector $IV_{AEAD}$. 
The session variables are combined into node-specific payloads: $P_A = K_s \parallel N_A \parallel N_S$ and $P_B = K_s \parallel N_S$.

\item \textbf{AEAD encryption:} Each payload is encrypted and authenticated using Ascon-128~\cite{dobraunig2021ascon} as \\$\text{AEAD\_Enc}(K, N, P, AD)$, where $K$ is the PUF response, $N = IV_{AEAD}$, $P$ is the payload, and $AD$ is the associated data:
\begin{align*}
(E_A, Tag_A) &= \text{AEAD\_Enc}(R_A,\ IV_{AEAD},\ P_A,\ ID_A \parallel ID_B \parallel T_S) \\
(E_B, Tag_B) &= \text{AEAD\_Enc}(R_B,\ IV_{AEAD},\ P_B,\ ID_B \parallel ID_A \parallel T_S)
\end{align*}
We choose Ascon as it is the National Institute of Standards and Technology's (NIST) choice for lightweight encryption \cite{turan2023status}. 
AEAD mode is used to provide integrity and authentication along with confidentiality~\cite{rogawayAEAD}.
The associated data binds both node identities and the server timestamp, so a modified timestamp or a substituted peer fails the tag.

\item \textbf{Server reply:} The server sends each node its encrypted payload ($E_x$), authentication tag ($Tag_x$), challenge $C_x$, the public $IV_{AEAD}$, and the AD Tag $Tag_x$.

\item \textbf{PUF response extraction and decryption (both nodes):} Each node first evaluates the PUF to reconstruct $R'_x$, and uses it to authenticate and decrypt its payload. 
If $Tag_x$ verification fails, the node aborts immediately, preventing chosen-ciphertext attacks and timestamp manipulation. 
Upon successful decryption, the node parses $K_s$, $N_S$, and $N_A$ from the payload and verifies $T_S$ and $N_A$ to confirm message freshness.

\item \textbf{Mutual authentication:} Both nodes now hold the same session key $K_s$. 
Each node computes a Hash-based Message Authentication Code (HMAC) \cite{krawczyk1997hmac} over a session-bound message $J_x = ID_x \parallel N_S \parallel T_S$ keyed with $K_s$ and sends it to the peer. 
Each node then independently recomputes the peer's expected HMAC and compares it against the received value. 
Matching HMACs confirm mutual authentication, session freshness, and shared possession of $K_s$.
Any secure MAC suffices, since an adversary without $K_s$ cannot forge a valid tag; we use the keyed BLAKE2s hash for its lightweight nature.

\item \autoref{fig:protocol} shows the complete protocol pictorially for reference.

\end{itemize}

\section{Security analysis and Design Decisions}
\label{sec:securityanalysis}
In this section, security analysis of DUFT and the overall protocol will be provided. 
Design decisions will be explained and supported by security analysis. 
First, we create a formal model of the adversary and their capabilities. 
Capabilities of the adversary are explained by the different physical layers of the design:
\subsection{Adversary Model}
\label{subsec:adversary_model}
\begin{itemize}
    \item Network Adversary (Dolev-Yao Model): The communication channel between IoT nodes and the server is assumed completely insecure under the Dolev-Yao threat model~\cite{dolev1983security}: an active adversary who can eavesdrop, intercept, inject, modify, and replay messages, but cannot compromise the underlying cryptographic primitives such as the Ascon cipher and HMAC functions.
    \item Physical Adversary: The adversary can capture the IoT device and extract non-volatile memory, but lacks micro-probing capability to read internal SRAM cells at power-up or to query individual SRAM interfaces, but may query as many inputs of the enlarged DUFT input space as they wish.
    \item Algorithmic Adversary: The adversary can intercept all helper data and challenges issued to a device, and may use this together with their physical-adversary capabilities to exploit anything leaked by the public helper data.
\end{itemize}
\subsection{Security Analysis of DUFT}
\subsubsection{CRP-Set, Physical Source, and Space Complexity}
\label{sec:crp-space-complexity}

Here we will establish our claims of the logical input-space of DUFT and the entropy of the physical source it wraps around.
We first distinguish these quantities to avoid overstating the security contribution of the logical input-space, and we make explicit the composition of the physical source, since this directly affects adversarial characterization cost.
\paragraph{Physical source:}
Each of the three SRAM chips is a fixed, silicon-realized function established at fabrication:
\begin{equation}
    \mathrm{SRAM}_i : \{0,1\}^{13} \rightarrow \{0,1\}^{32},
    \qquad i \in \{1,2,3\},
    \label{eq:single-sram-source}
\end{equation}
with a domain of only $2^{13} = 8{,}192$ addressable physical locations each.
Because the combined address $A = a_3 \Vert a_2 \Vert a_1$ and combined output $D = d_3 \Vert d_2 \Vert d_1$ are formed by concatenation of architecturally independent per-chip addresses~\eqref{eq:address_concatenation}--~\eqref{eq:powerup_data_concatenation}, the combined function
\begin{equation}
    \mathrm{MSRAM} : \{0,1\}^{39} \rightarrow \{0,1\}^{96}
    \label{eq:msram-source}
\end{equation}
is not a 39-bit physically-realized function but a deterministic recombination of three separately enumerable $2^{13}$ physical SRAM sources.
The cost for a physical adversary to fully characterize $\mathrm{MSRAM}$ is therefore
\begin{equation}
    n_{\mathrm{reads}} = \sum_{i=1}^{3} 2^{13} = 3 \times 2^{13}
    = 24{,}576
    \label{eq:char-cost}
\end{equation}
reads, after which $\mathrm{MSRAM}(A)$ can be computed offline for any of the $2^{39}$ combined addresses, and by extension the DUFT output for any challenge, since each is a deterministic function of this fully-known map composed with public operations.
Repeated reads at this cost also recover the stability mask table and, through it, the corrected array and hence $K_{dev}$, so an adversary with address-level access gets the traversal schedule and the round data alike; no part of the construction is hidden from him.
Our security claims in this section therefore hold only under the physical adversary assumption of~\autoref{subsec:adversary_model}, which excludes direct, address-level query access to the SRAMs, and DUFT does not raise this characterization cost since it operates entirely as a wrapper of the physical source.

The physical source, its combination, and hence the logical challenge space depend on the SRAM source(s) chosen, how many are placed in parallel, and the traversal seed size. DUFT is flexible in all three, and we use three $8192\times32$-bit SRAMs with a 64-bit seed.
Any configuration must observe the caveat above, that the characterization cost for an adversary with SRAM-interface access is $n_{reads}$ for that setup.
We demonstrate in~\autoref{sec:Methodology} an MCU with embedded SRAM that does not expose the SRAM interface. An ASIC implementation would likewise not expose it, and we leave that to future work.

\paragraph{DUFT Input Space:}
DUFT's public input is $C = \nu \Vert A$.
The MSRAM address $A$ chooses the starting point of the traversal, and the traversal seed $\nu$ determines the feedback traversal for that address. Since the power-up stable and unstable bit positions of each SRAM are unique, so is the traversal for a given $\nu$ and $A$.
Without the fingerprint of the SRAMs included in the traversal, the traversal for any underlying SRAM physical source would be the exact same, and without $K_{dev}$ it would be recomputable by anyone holding the public mask table.
The input-space is therefore
\begin{equation}
    |\mathcal{C}| = 2^{w_\nu + w_A} = 2^{64+39} = 2^{103}.
    \label{eq:logical-challenge-space}
\end{equation}
Each $C \in \mathcal{C}$ deterministically selects a $n_r$-round traversal path over the fixed physical source of~\eqref{eq:msram-source} via the BLAKE2s feedback of~\eqref{eq:roundhash_counter}.
Distinct challenges are computationally decorrelated traversal paths over a common, small physical object, not independent physical secrets: $|\mathcal{C}| = 2^{103}$ characterizes the input space to the DUFT wrapper, not the physical entropy of the underlying SRAM source.

\paragraph{Enrolled CRP set:}
The server enrolls and stores a practical subset of size $n_{CRP}$ (such as $n_{CRP} = 10{,}000$ per device) of the logical challenge space, together with each entry's response and helper data $(R, B)$.
The logical challenge space is equivalent to the input-space of DUFT.
Once exhausted, a new set is enrolled which excludes all previously enrolled sets, and all CRPs within a set are used only once, so DUFT's $2^{103}$ input-space provides a supply that can be refreshed indefinitely without the server storing a large database. This is a standard practice for PUF-based authentication.
However, the entropy of the underlying source being drawn on is still much smaller than the input-space, so this CRP supply is dependent on that underlying entropy not being compromised, which we address by obfuscating the raw MSRAM outputs with a final BLAKE2s hash.

\subsubsection{Entropy and Leakage Definitions}
\label{subsubsec:entropy-definitions}

Before stating our formal security claim, we define the quantities used throughout this section, distinguishing what is measured empirically (Section~\ref{sec:Methodology}) from what is assumed.

\paragraph{Population min-entropy ($e_{bits}^{pop}$).}
We define the per-bit min-entropy of the physical SRAM source as
\begin{equation}
    e_{bits}^{pop} = -\log_2\!\left(p_{max}\right),
    \label{eq:min-entropy-def}
\end{equation}
where $p_{max}$ is the probability of the most likely value at a given bit position, estimated empirically across the evaluated chip population; $2^{-e_{bits}^{pop}}$ is thus the probability that an adversary who has profiled this SRAM design at the population level, but not read the target device, correctly guesses that bit on an unseen chip.

\paragraph{Sequential min-entropy ($e_{bits}^{seq}$).}
Whereas $e_{bits}^{pop}$ characterizes cross-device predictability at a fixed position, $e_{bits}^{seq}$ characterizes within-device predictability, capturing statistical dependence between neighbouring bits that $e_{bits}^{pop}$ cannot detect.
Following NIST SP 800-90B~\cite{nist80090b}, we define the min-entropy of a non-Independent and Identically Distributed (non-IID) sequence of stable bits read in physical order from a single device as
\begin{equation}
    e_{bits}^{seq} = -\log_2\!\left(p_{\max}^{\,\text{seq}}\right),
    \label{eq:h-original-def}
\end{equation}
where $p_{\max}^{\,\text{seq}}$ is the highest achievable success probability of any predictor $\pi$ guessing $X_i$ given the preceding bits $(X_1,\dots,X_{i-1})$:
\begin{equation}
    p_{\max}^{\,\text{seq}} = \max_{\pi} \; \Pr\big[\pi(X_1,\dots,X_{i-1}) = X_i\big].
    \label{eq:h-original-pmax}
\end{equation}
NIST SP 800-90B's non-IID track estimates $p_{\max}^{\,\text{seq}}$ empirically using ten estimators spanning distinct prediction strategies (listed in~\autoref{tab:nist90b}), reporting the minimum across all ten on the principle that an adversary need only exploit the single most effective strategy:
\begin{equation}
    e_{bits}^{seq} = \min_{k=1,\dots,10} e_{bits}^{seq}[k],
    \label{eq:h-original-min}
\end{equation}
where $e_{bits}^{seq}[k]$ is the estimate from estimator $k$. Unlike $e_{bits}^{pop}$ this requires no independence assumption between positions, and directly bounds an adversary exploiting spatial structure within a single device's stable-bit readout.

\paragraph{Helper-data leakage ($l_m$, $l_{BCH}$).}
Our helper data leaks two quantities, and they are measured in different units.
The stability mask leaks the positions of the stable and unstable bits in the pre-hash response $R_{SRAM}$~\eqref{eq:sramresponse}, and leaks that the unstable positions are set to 0, which gives those positions zero entropy.
This is a count of bit positions, since it removes them from the budget outright rather than degrading them.
In~\autoref{subsubsec:error-corr-experiments} we experimentally determine the number of unstable bits in a round MSRAM read $D_r$, under high temperature testing, to average 42 for our reference hardware:
\begin{equation}
    l_m = 42, \qquad s = w_D - l_m = 54,
\end{equation}
where $s$ denotes the number of retained stable bits per round read.
We assume with empirical evidence that no information about the logical values of stable bits beyond their positions is leaked, and mask positions across enrolled CRPs do not correlate with device physical layout or with one another, as validated by our structural-leakage experiment (\autoref{subsubsec:error-corr-experiments}).

The window sketches, by contrast, leak entropy equal to their parity length, per the worst-case bound of Dodis et al.~\cite{dodis2004fuzzy}, and this bound is on \emph{entropy}, not on bit positions:
\begin{equation}
    \tilde{\mathbf{H}}_\infty(G_{arr} \mid P_{tab}) \;\geq\; \mathbf{H}_\infty(G_{arr}) - l_{BCH},
    \qquad l_{BCH} = n_{win} \cdot w_f \cdot t,
    \label{eq:dodis-bound}
\end{equation}
where $H_{\infty}(X)$ is the min-entropy of $X$, $G_{arr}$ is the golden reference of the whole array and $n_{win}$ the number of windows.
This is a single fixed deduction against the array, taken once at enrollment, and not a per-challenge quantity.
Charging it to the traversal requires attributing it across the array's stable bits.
We do this proportionally, which is the conservative reading, since we cannot control which bits an adversary would prefer the leak to fall on:
\begin{equation}
    e_{bits}^{\mathrm{eff}} = e_{bits}^{\mathrm{final}} - \frac{l_{BCH}}{S_{arr}},
    \label{eq:eff-entropy}
\end{equation}
where $S_{arr}$ is the total number of stable bits in the array.
For our configuration $S_{arr} = 442{,}368$, and $l_{BCH}$ is $16{,}560$ bits on the Cypress CY62256 and $38{,}160$ on the STM32F407, so the deduction is $0.037$ and $0.086$ bits per stable bit respectively.
\paragraph{Round and traversal min-entropy ($e_D$, $e_R$).}
$e_D$ denotes the min-entropy contributed by a single round's MSRAM output $D_r$, over its retained stable positions only:
\begin{equation}
    e_D = s \cdot e_{bits}^{\mathrm{eff}},
    \label{eq:round-entropy}
\end{equation}
where $s$ is the number of stable bits in $D_r$.
$e_R$ denotes the min-entropy of the accumulated pre-hash response:
\begin{equation}
    e_R = n_r \cdot e_D.
    \label{eq:traversal-entropy}
\end{equation}
There is no further deduction at this point, since the sketch leakage has already been charged into $e_{bits}^{\mathrm{eff}}$ by~\eqref{eq:eff-entropy}.
We establish $e_{bits}^{\mathrm{final}}$ by combining $e_{bits}^{pop}$ and $e_{bits}^{seq}$ in~\autoref{subsec:respsecurity}.

\subsubsection{Security of Response}
\label{subsec:respsecurity}
First we state our assumptions to support our claim:

\paragraph{Assumption 1 (Within-challenge round independence).}
The 96-bit MSRAM outputs $D_r$ at each round $r \in \{0,\dots,n_r-1\}$ within a single challenge's traversal are independent, so their min-entropy contributions accumulate additively across rounds.
Address distinctness across rounds is guaranteed by construction (Algorithm~\ref{alg:rtcr}, lines~\ref{line:feedback}--\ref{line:retryloop2}), and~\eqref{eq:roundhash_counter}, which rejects and re-derives any next-round address already visited earlier in the traversal.
We do not assume independence across distinct enrolled challenges, addressed in the remark following Theorem~\ref{thm:rtcr-security}. 

\paragraph{Assumption 2 (Established source entropy).}
We take the per-bit population min-entropy $e_{bits}^{pop}$ and sequential min-entropy $e_{bits}^{seq}$ (\autoref{subsubsec:entropy-definitions}) as empirically established for our tested hardware (\autoref{sec:Methodology}), and use $e_{bits}^{\mathrm{final}} = \min(e_{bits}^{pop}, e_{bits}^{seq})$ as our working per-bit figure.
Rigorously establishing these quantities across process variation, manufacturing lots, and environmental conditions is a substantial characterization problem outside our scope. This dependency is not unique to DUFT, since any construction built atop an SRAM source, including hashing a single raw read, can only claim a given security level if the source's entropy has first been established to support it.

\paragraph{Assumption 3 (Bounded helper-data leakage).} 
The stability mask $M$ leaks no information about stable-bit values beyond their positions, and mask positions do not correlate with device layout or with one another, as validated by our stability-mask experiment (\autoref{subsubsec:puf_metrics}).
Leakage is bounded by $l_m = 42$ masked positions per round (measured average unstable bits under temperature variation, \autoref{subsubsec:error-corr-experiments}), leaving $s = 54$ retained positions, and by $l_{BCH} = n_{win} w_f t$ bits of min-entropy against the array as a whole, charged into $e_{bits}^{\mathrm{eff}}$ by~\eqref{eq:eff-entropy} for the parity bits.
We additionally assume both tables are public. Although held in device non-volatile memory, they may be recovered by an adversary who captures and dumps the device, and our analysis grants this without qualification.
$K_{dev}$ is not public, but we do not credit it with any min-entropy either. An adversary who somehow learns it recovers the address schedule and nothing else, which returns the construction to the public-schedule case and leaves~\eqref{eq:security-bound} unchanged.

\paragraph{Assumption 4 (BLAKE2s as a random oracle / PRF).}
Unkeyed BLAKE2s as used in our construction~\eqref{eq:blakehash}, the device-key derivation of~\eqref{eq:devkey}, and the round feedback hashes of~\eqref{eq:roundhash} and~\eqref{eq:roundhash_counter} behaves as a random oracle for round-address derivation and as a pseudorandom function for privacy amplification. 
This is a standard modeling assumption for hash-based fuzzy extractor constructions~\cite{dodis2004fuzzy}.
No structural distinguisher against BLAKE2s is currently known~\cite{aumasson2013blake2}.

\begin{theorem}[DUFT response security]
\label{thm:rtcr-security}
Under Assumptions 1--4 above, for any adversary $\mathcal{A}$ confined to the physical adversary model of ~\autoref{subsec:adversary_model} (no direct SRAM address query access), the probability that $\mathcal{A}$ correctly predicts the DUFT response $R$ for a given challenge $C$, without possessing the physical device, is bounded by
\begin{equation}
    \Pr[\mathcal{A} \text{ predicts } R \mid C]
    \;\leq\;
    2^{-e_R}
    \;+\;
    \mathrm{Adv}^{\mathrm{PRF}}_{\mathrm{BLAKE2s}}(\mathcal{A}),
    \qquad
    e_R = n_r \cdot s \cdot e_{bits}^{\mathrm{eff}},
    \label{eq:security-bound}
\end{equation}
where $s = w_D - l_m$ is the number of stable bits retained per round, $n_r$ is the number of DUFT rounds, $e_{bits}^{\mathrm{eff}}$ carries the array-wide sketch deduction of~\eqref{eq:eff-entropy} under Assumptions 2 and 3, and $\mathrm{Adv}^{\mathrm{PRF}}_{\mathrm{BLAKE2s}}(\mathcal{A})$ is $\mathcal{A}$'s advantage in distinguishing BLAKE2s from a random function (Assumption 4).
\end{theorem}

\begin{proof}
By Assumption 1, the min-entropy contributions of the $n_r$ rounds accumulate additively, $\sum_{r=0}^{n_r-1} e_D = n_r \cdot e_D$.
By Assumption 3 and the bound of Dodis et al.~\eqref{eq:dodis-bound}, the parity tables reduce the conditional min-entropy of the array by at most their total length, and~\eqref{eq:eff-entropy} charges this to each stable bit before the traversal draws on it.
Since the parity is fixed per device rather than issued per challenge, this deduction does not grow with the number of enrolled CRPs, and guessing the accumulated pre-hash response succeeds with probability at most $2^{-e_R}$.

The adversary cannot do better by attacking the rounds separately.
Under Assumption 4 the feedback hash is a random oracle, and none of its inputs $K_{dev}$, $\nu$, $r$, $ctr_r$ or $m_r$ is a function of any $D_r$, so the chain neither reveals a round's data nor gives a relation against which to test a candidate for one.
$K_{dev}$ is a hash of the whole array rather than of the traversal, so it carries no per-round relation either, and since it is derived from the same source as the response it contributes nothing to $e_R$. 
Its role is to withhold the schedule, not to supply entropy.
Guessing $K_{dev}$ which is a 128-bit digest is equivalent to guessing a 128-bit number or the pre-image which is the complete three combined arrays.
The final BLAKE2s hash provides privacy amplification, so the only route left to predicting $R$ is guessing $R_{SRAM}$ outright, bounded by $2^{-e_R}$.
The term $\mathrm{Adv}^{\mathrm{PRF}}_{\mathrm{BLAKE2s}}(\mathcal{A})$ covers the possibility, believed negligible but unproven, that BLAKE2s departs from this idealized behavior.
\end{proof}

\paragraph{Remark.}
With $n_r = 7$ on both parts and $e_D \approx 18.5, 19.7$ bits per round (computed from $e_{bits}^{\mathrm{eff}}$) for our two reference experiments in~\autoref{sec:Methodology}, the combinatorial term of Eq.~\eqref{eq:security-bound} evaluates to below $2^{-128}$.
We report this as an engineering security estimate under the stated assumptions, not an unconditional cryptographic proof, conditional on the physical source satisfying Assumptions 1--4.
This dependency is not unique to DUFT. No overlying wrapper can manufacture entropy absent from its input physical source, so any construction over an SRAM source, including hashing a single raw read~\cite{Gassend2002Controlled}, can claim $n$ bits only if $n$ bits of extractable min-entropy are first established.
DUFT utilizes the estimated or established entropy of the source to derivce the number of rounds via
\begin{equation}
    n_r = \left\lceil \frac{n}{s \cdot e_{bits}^{\mathrm{eff}}} \right\rceil, \qquad
    e_{bits}^{\mathrm{eff}} = e_{bits}^{final} - \frac{n_{win} w_f t}{S_{arr}}, \qquad
    t = \left\lceil \mu + c\sigma \right\rceil,
    \label{eq:nr-conditional}
\end{equation}
where $n$ is the target security level, $s = w_D - l_m$ is the measured number of stable bits retained per round, $e_{bits}^{final}$ is the true per-bit min-entropy of the source, and $\mu, \sigma$ are the mean and standard deviation of the window error count.
$w_f$ is the smallest field degree with $2^{w_f} - 1 \geq win \cdot w_D + w_f t$, and the window size $win$ is the free parameter: larger windows raise $e_{bits}^{\mathrm{eff}}$ by spending less of each block on parity, but decoding cost $O(n \cdot t)$ becomes impractical well before the array size is reached.
Practitioners should recompute all of this on their own devices rather than carry over our reference values.
Entropy estimation must be performed as rigorously as possible or a established entropy-source can be used to have certainty of the security of the response, and error-correction must be tested thoroughly to avoid fatal correction failures.

\subsection{Design of the Address Schedule}
\label{subsec:feedbackdesign}
We consider three options to drive the address traversal and compare them.
The first option is to use nothing secret at all: build the schedule from just the public seed and round number, with or without the stability mask.
But the stability mask only depends on a $w_a$-bit sub-address $a$, and the same address always gives the same mask.
That means the full set of possible masks has only $n_{SRAM} \cdot 2^{w_a} = 24{,}576$ entries.
So an adversary can rebuild this schedule by reading the mask table off a captured device.
Although this would only give him the schedule for that particular SRAMs since each instance still walks its own unique path.
Therefore, the stability mask only provides a per-challenge and per-chip unique traversal but the schedule is public. 

The second option is to use the round data $D_r$ in place of $m_r$, which actually is secret while also unique per-challenge and per-chip.
We do not do this because if any part of the schedule is ever observed, the attacker knows every term in the equation for $A_{r+1}$ except $D_r$.
He can then guess $D_r$ for a single round, check that guess by recomputing the hash and comparing it to the observed address, and repeat this round by round for a peeling attack.
That drops the cost of recovering a full response from $2^{e_R}$ down to about $n_r \cdot 2^{e_D}$.
We do not want a scheme that falls apart that fatally from a traversal schedule leak.

So we go with the third option: we still use the stability mask, seed and round number, but also a secret hash of the complete corrected array of the three SRAMs $K_{dev}$.
If we only use $K_{dev}$ and not $m_r$, the feedback traversal would not be unique across challenges apart from the first address. 
Since $K_{dev}$ is a hash of the corrected stable bit array of the three SRAMs, guessing its pre-image as difficult as guessing the three complete SRAM arrays.
$K_{dev}$ is not an on-device secret, but rather a function of our live PUF instance at power-up.
An attacker who only has $M_{tab}$ and $P_{tab}$ can not compute it, and so can not compute the schedule either.
$K_{dev}$ is computed only once per power-up in volatile memory and then lost on powering off.
In case the $K_{dev}$ is leaked, the three SRAM arrays that are the pre-image of the hash are still safe because of Blake2s's pre-image resistance.
The leak only compromises the address traversal schedule for that particular MSRAM instance.
If the schedule can be recovered, an attacker can tell which rounds of two different enrolled responses land on the same address, and therefore which bits of those two responses are identical before hashing.
Although, he only learns the positions, and not the values, it does mean that breaking one response also hands him the shared bits of every response that overlaps with it.
With $K_{dev}$ in the chain, he can not tell which responses overlap, and he can not cherry-pick a challenge based on partial knowledge.
Thisc stops a partial break from spreading to other responses.

Using $K_{dev}$ we create an address traversal that is secret using the PUF array fingerprint, and using $m_r$ or the stability mask fingerpint we make the traversal unique to that particular PUF array.
This traversal extracts us a fractional entropy of that array which is our complete novel design. 

\subsection{Security Analysis of the Protocol}
\label{subsec:protocolsecurity}
The protocol relies on at least 128 bits of min-entropy from the PUF, using the assumptions and measurements from~\autoref{subsec:respsecurity}, plus an effectively unlimited supply of single-use CRPs from the DUFT input space described in~\autoref{sec:crp-space-complexity}.
The server generates the session key $K_s$ and both nonces, so neither node has to generate any keys itself — each node just does one PUF evaluation, one Ascon decryption, and two hashes.
Using the 128-bit PUF responses ($R_A, R_B$) to encrypt the session key is the core idea behind the protocol's security: the raw response is never sent over the network, so a Dolev-Yao network adversary can't just grab it off the wire.
We now go through what an adversary can actually get out of each attack our threat model allows.

\paragraph{\textbf{Eavesdropping and impersonation.}}
A passive adversary watching the network sees $C_x$, $E_x$, $Tag_x$, $IV_{AEAD}$, $T_A$, and the two HMACs, but never the PUF response itself.
Since the payload is encrypted under $R_x$, recovering the session key $K_s$ means recovering $R_x$ first, and that is exactly the guessing problem bounded by Theorem~\ref{thm:rtcr-security}: it is computationally infeasible without the physical device in hand.
Decrypting the payload only ever happens inside the device's volatile memory during a live PUF evaluation, and the response is never written to storage.
Watching many sessions does not help either, since every response is hashed before it is used and each CRP is only ever spent once.

\paragraph{\textbf{Message tampering.}}
If an attacker changes $E_x$ or $Tag_x$ in transit, the AEAD tag check on the receiving end fails immediately, and forging a tag that passes requires knowing $R_x$, which the attacker does not have.
Tampering with the challenge $C_x$ instead just sends the node down a different traversal of the array, producing a different response that can not decrypt the payload and again fails the tag check.

\paragraph{\textbf{Replay.}}
NodeA's nonce $N_A$ is echoed back inside its own encrypted payload, so NodeA can check it is talking to a live session and not a replay: an old server reply would carry the wrong nonce and get rejected outright.
The timestamp $T_S$ also puts a ceiling on how long any intercepted message stays valid.
Because the authentication message $J$ is tied to the fresh session nonce $N_S$ and timestamp $T_S$ rather than to fixed identifiers, an HMAC captured from an earlier session simply will not verify in a new one.
And since CRPs are single-use, replaying an old $C_x$ does not work either, because the server has already retired it.

\paragraph{\textbf{Man in the middle.}}
For an attacker to sit in the middle, he would need to hold a separate valid session with each node while both nodes think they are talking directly to each other.
That requires forging a valid $(E_x, Tag_x)$ pair under $R_A$ for NodeA and under $R_B$ for NodeB, and he has neither response, so at best he can only relay messages unchanged, which makes him nothing more than a wire.
The one attack worth calling out separately is peer substitution: since the very first message is sent unencrypted, an attacker could swap in $ID_B$ for a device he controls.
But the intended peer's identity is attached into each node's associated data for the AEAD tag, so substituting the peer causes the tag check to fail.

\paragraph{\textbf{Physical capture.}}
If an attacker physically captures a node, they do not get anything useful right away: the mask and parity tables stored on the device are already public, and the device key and the response only ever exist in volatile memory during a live evaluation, so a powered-down device gives up neither the array contents nor the address schedule.
The picture is different at the server, though.
The server's database stores each response $R$ in the clear (Algorithm~\ref{alg:enrollcrp}), so if the server itself is ever compromised, every session key tied to a CRP that hasn't been used yet is exposed all at once, without the attacker needing to touch a single device.
Protecting that server-side storage is outside the scope of this paper, but we acknowledge that it is not addressed in this design.
\section{Experimental Setup and Methodology}
\label{sec:Methodology}
\subsection{Hardware}
For implementing DUFT, we focus on two SRAMs previously used in literature.
\begin{itemize}
    \item The Cypress CY62256NLL~\cite{cypresscy62256} which has been thoroughly evaluated by Sajim et al.~\cite{Sajim2018cypress}.
    \item The STM32F407 chip~\cite{stm32f407_datasheet} having an embedded SRAM. This SRAM has been evaluated by Barbareschi et al.~\cite{barbareschi2015testingstm32}
\end{itemize}
The Cypress SRAM is a 256-Kbit SRAM chip with a data width of 8-bits and an address width of 15-bits. 
The SRAM provides parallel address and data interfaces to the SRAM array. 
For our PUF use case, we are interested in just reading the power-up values of the SRAM. 
To use it as a 13-bit address and 32-bit data SRAM, we read 4 bytes in a burst. 
The 13 upper-bits of the 15-bit address specify the 32-bit word to be read from the array and by cycling 0-3 in the lower 2-bits of the address, we read out the 32-bits of data.
Our Micro-Controller Unit (MCU) of choice to interface with this chip for a single node was the Raspberry Pi 5 (RPi5) or the Raspberry Pi500 (RPi500)\cite{raspberrypi5}. 
Three of these SRAMs are connected to the RPi5/500 through General Purpose Input and Output Pins (GPIOs) and are also powered up through a GPIO pin to support powering up and powering down the SRAMs on the fly.
The Cypress implementation will feature DUFT and the full end-to-end protocol implementation.

\begin{figure}[t]
\centering
 \includegraphics[width=\columnwidth]{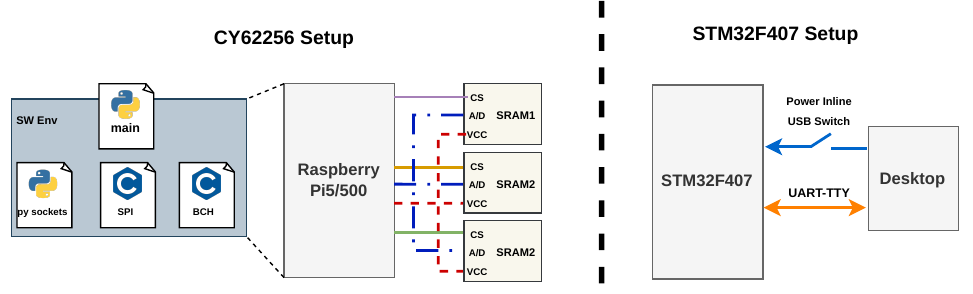}
\caption{Hardware setup of RPi5/500 with the Cypress SRAMs and the STM32F407 embedded SRAM for evaluation.}
\Description{Hardware setup for the Cypress SRAMs with three SRAMs interfaced to the Raspberry Pi. The STM32F407 with embedded SRAM is interfaced with desktop via UART and an inline-switch USB cable for power-cycling.}
\label{fig:hw_setup}
\end{figure}

We use the STM32F407-DISC1 board to demonstrate DUFT on embedded SRAM with a lightweight MCU implementation, distinct from our discrete Cypress SRAM setup. 
We cannot confirm this board satisfies our physical adversary model (\autoref{subsec:adversary_model}) without qualification: voltage fault injection has been demonstrated against the STM32F407 specifically, capable of downgrading readout protection and re-enabling debug access, and our firmware evaluates the response in software after boot, so a debug compromise could expose intermediate values in RAM.
We therefore treat this board as a proof of concept for DUFT on embedded SRAM rather than as satisfying the stated model; a production deployment requires readout protection enabled, and debug disabled.
We demonstrate only the DUFT mechanism on this board, not the end-to-end protocol.
Of the 192KB of SRAM the STM32F407 ships with, we use 64KB of the the Core-Coupled Memory (CCM) SRAM and 32KB of the System SRAM for a total of 96KB of SRAM array to be used as our PUF.
We split this 96KB array into three chunks of 32KB which we will be treating like three parallel SRAMs as our MSRAM.
We could have split into two-parts or use just one contiguous array as DUFT is flexible but this is done for better consistency in demonstration.
The per-address stability mask table occupies 96\,KB, held in on-board flash and comfortably within the 1\,MB the part provides.
\autoref{fig:hw_setup} shows the hardware setup for both SRAMs.

For the Cypress setup for experimenting and evaluating the protocol, two such RPi5/500 with 3 SRAMs connected via GPIO are used as nodes of the protocol. 
A desktop computer is used to act as the trusted third-party server maintaining the CRP database and key generation. 
The server and the two nodes are connected to the same LAN via Ethernet. 
Ethernet will be the mode of communication for the protocol.
\subsection{Software}
For Cypress, the socket-programming ethernet communication is implemented in Python, chosen for its community support on the RPi5/500 and for ease of implementation, while the SRAM reads, feedback, DUFT derivation and the BCH error correction are implemented in C.
The desktop PC acting as the TTP server runs a multi-threaded script managing sockets to several nodes, and maintains a JSON database of enrolled NodeIDs and their CRPs.
A packet-based approach is used, where the requester, whether server or node, sends a request packet carrying a command for the action being requested.
The server can request enrollment from a node, and a node can request authentication and key establishment with another node.
The Python Ascon library encrypts the shared session key with each node's response, and the nodes use the same library to decrypt it.
Performance is evaluated with the Python time library, using \textit{perf\_counter\_ns()} for wall-clock time and \textit{process\_time\_ns()} for CPU time.

For the STM32F407, we use a linker script to reserve the CCM and a portion of System RAM as our PUF memory and then instantiate them as 32-bit word arrays in the main code.
All code is implemented in C.
We implement a wrapper function to utilize the three 32KB chunks of the PUF array as the MSRAM.
We use Blake2s with 128-bit digest from its standard github implementation.
The mask table is placed in a dedicated flash section along with the parity table.
We use an inline-button USB cable to power-cycle the board, as unlike Cypress we can not power-cycle via GPIO through software.
Performance is evaluated with the Data Watchpoint and Trace (DWT) cycle counter while running the board on a 168MHz clock.

\begin{table}[htbp]
\centering
\small
\setlength{\tabcolsep}{4pt}
\caption{SRAM PUF quality metrics.}
\label{tab:puf_metrics}
\renewcommand{\arraystretch}{1.2}
\begin{tabular}{lcccc}
\hline
\textbf{Metric} & \textbf{CY62256} & \textbf{STM32F407} & \textbf{Ideal} & \textbf{Unit} \\
\hline
Inter-chip HD               & 41.03 & 42.23  & 50.00  & \% \\
Intra-chip HD               & 2.50  & 3.72  & 0.00   & \% \\
Uniformity                  & 53.32 & 48.34  & 50.00  & \% \\
Bit-aliasing                & 48.22 & 49.01 & 50.00  & \% \\
Population Min-entropy (per bit)       & 0.41 & 0.45   & 1.0000 & bits \\
Stability Maks Inter-chip HD       & 27.66 & 29.23   & 50.00 & \% \\
Unstable Bits Overlap       & 4.23 & 3.22   & 0.00 & \% \\
\hline
\end{tabular}
\end{table}
\subsection{Performance Evaluation and Results}
\subsubsection{PUF Metrics}
\label{subsubsec:puf_metrics}
We now evaluate the PUF metrics for both SRAMs, since they are relevant to design decisions and security analysis.
Table~\ref{tab:puf_metrics} reports inter-chip and intra-chip Hamming distance, uniformity, bit-aliasing, population min-entropy, stability mask Inter-Chip HD and unstable bits overlap collected over thirty-two (32) Cypress chips and twenty-eight (28) STM32F407-DISC1 boards.

Our uniformity and per-bit min-entropy results fall short of an ideal PUF, which we attribute to a poor sample-set with common die variations, since~\cite{barbareschi2015testingstm32} reports near-ideal uniqueness for the STM32F407 and~\cite{korenda2019proof} uses the CY62256.
We test uniqueness and min-entropy only across stable bits, as unstable bits do not form part of the SRAM PUF.
Our stability-mask inter-HD is also low, because the majority of array bits are stable and stable bits are set to 1 in the mask. To confirm the masks are still unique, we measure the overlap of unstable-bit positions across our chip population, at $\approx$4\%, meaning only 4\% of all unstable positions are common to every chip.

We next run the NIST SP 800-90B test on both SRAMs to establish the sequential min-entropy $e_{bits}^{seq}$ (\autoref{subsubsec:entropy-definitions}). \autoref{tab:nist90b} shows the results.
For the CY62256, each SRAM is only 256 Kbit and so falls short of the test's recommended 1-million-bit sample, which may make the estimate less accurate.
Taking the minimum of the population and sequential min-entropy for each part gives $e_{bits}^{CY62256} = 0.38$ and $e_{bits}^{STM32F407} = 0.45$: engineering estimates from our chip-population sample, not a cryptographic proof.
A careful DUFT implementation will need underlying physical SRAM sources with established per-bit min-entropy.

\begin{table}[htbp]
\centering
\small
\setlength{\tabcolsep}{4pt}
\caption{NIST SP 800-90B non-IID entropy estimates (min-entropy per symbol, bits).}
\label{tab:nist90b}
\renewcommand{\arraystretch}{1.2}
\begin{tabular}{lcc}
\hline
\textbf{Estimator} & \textbf{STM32F407} & \textbf{CY62256} \\
\hline
Most Common Value (MCV)     & 0.953425 & 0.889053 \\
Collision                   & 0.501964 & 0.433997 \\
Markov                      & 0.751864 & 0.717122 \\
Compression                 & 0.518569 & 0.382970 \\
T-Tuple                     & 0.602691 & 0.431024 \\
Longest Repeated Substring (LRS) & 0.833183 & 0.757544 \\
Multi Most Common in Window (MultiMCW) & 0.839461 & 0.674053 \\
Lag Prediction               & 0.757289 & 0.732132 \\
Multi Markov Model with Counting (MultiMMC) & 0.693115 & 0.652300 \\
LZ78Y Prediction              & 0.757287 & 0.702062 \\
\hline
\textbf{$e_{bits}^{seq}$ (min across all estimators)} & \textbf{0.501964} & \textbf{0.382970} \\
\hline
\end{tabular}
\end{table}
\subsubsection{Error-correction}
\label{subsubsec:error-corr-experiments}
To support our robust error-correction claim, we perform four experiments:
\begin{itemize}
    \item Test SRAM reliability under varying-voltage conditions and extreme temperatures.
    \item Create a distribution of stable bits in 96-bit $D_r$ round read in high temperature conditions.
    \item Perform multiple reads of both SRAM arrays BCH window $win$ after stability mask application to characterize mean $\mu$ and standard deviation $\sigma$ of the number of errors per window.
    \item We test if every bit position in 32-bit reads across a single SRAM chip is equally likely to be unstable.
\end{itemize}

\begin{figure}[t]
\centering
 \includegraphics[width=0.8\columnwidth]{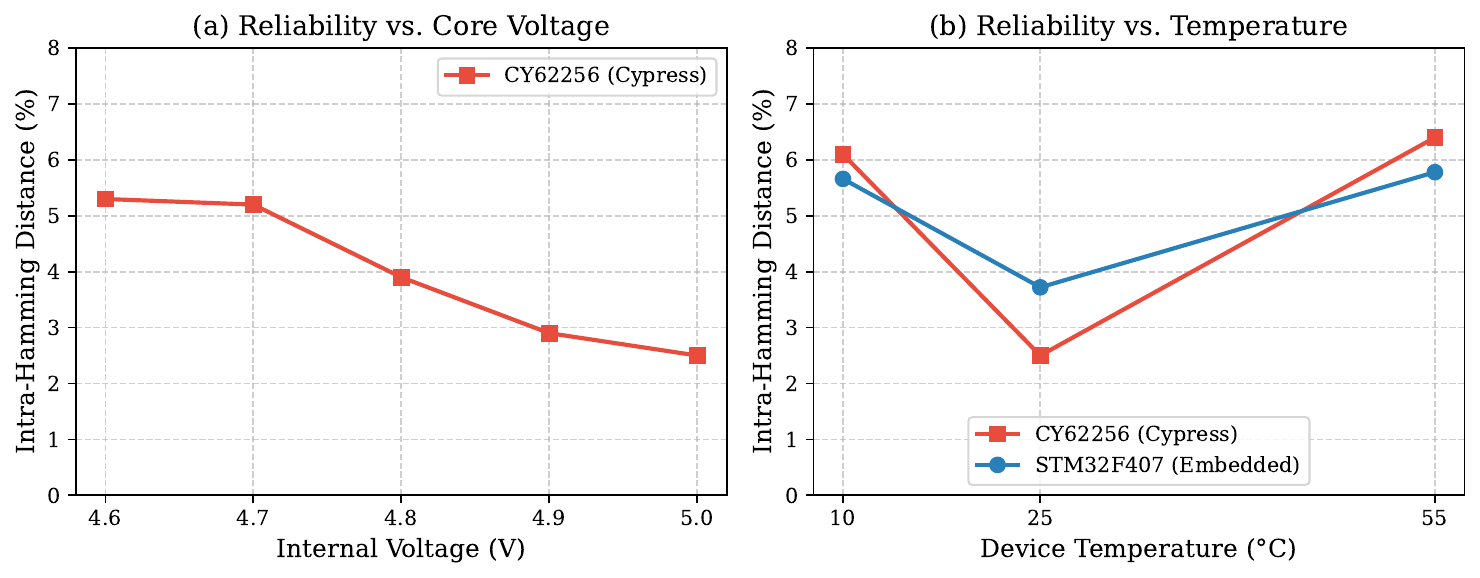}
\caption{Reliability evaluation of Cypress and STM32F407 SRAMs under varying environmental Conditions.}
\Description{Two graphs showing how the reliability of the our reference SRAMs vary with voltage and temperature respectively. The voltage graph only includes results for the Cypress SRAMs.}
\label{fig:tempvoltrely}
\end{figure}

Our varying-voltage and temperature experiments show intra-HD increasing slightly across our range of variation (\autoref{fig:tempvoltrely}). We were unable to sweep the full voltage and temperature range for both SRAMs in this paper to stay within scope, and we were only able to perform the voltage variation on the Cypress CY62256, but our results agree with prior work by~\cite{Sajim2018cypress} for CY62256 and Zeinzinger et al.~\cite{zeinzingerstm32temp} for the STM32F407 under similar variation.

\begin{figure}[t]
\centering
 \includegraphics[width=0.9\columnwidth]{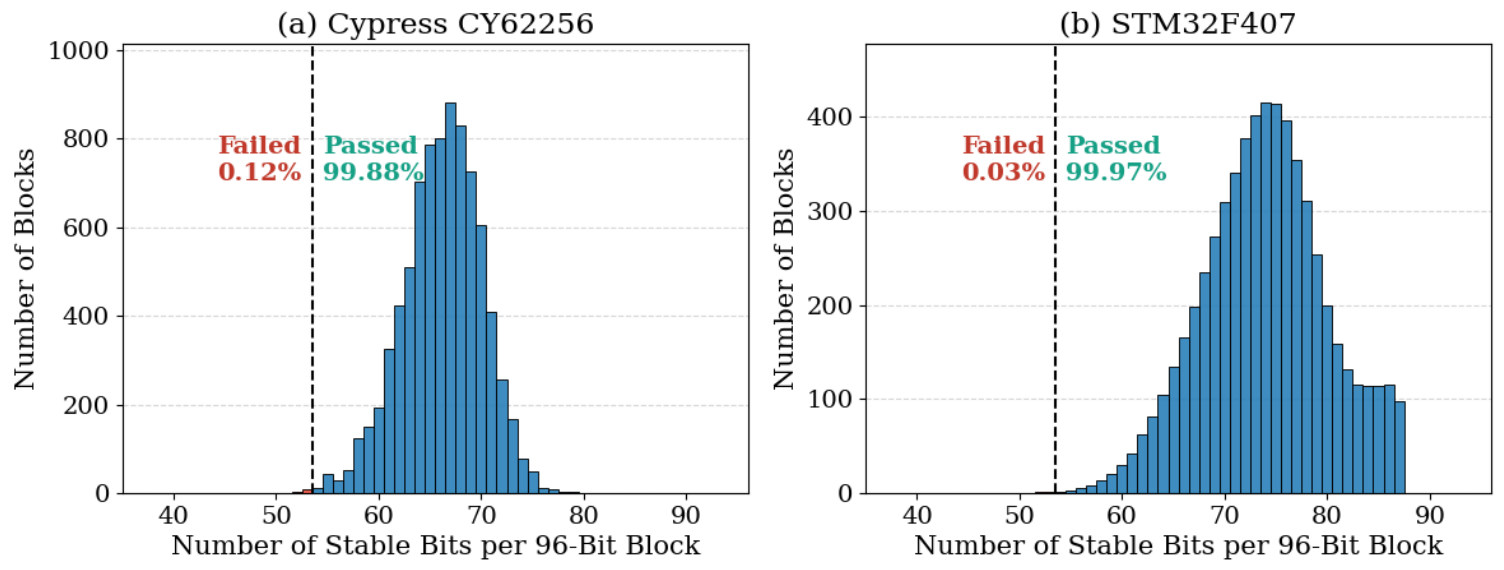}
\caption{Stable-bit distribution across 96-bit MSRAM blocks with the selected cutoff threshold.}
\Description{Two histograms that show the stable bit distribution for both of our reference SRAMs. They resemble a bell curve distribution with both having a minimum or approximately 54 stable bits.}
\label{fig:stability_bell}
\end{figure}

Since it is the number of unstable bits that increases with temperature~\cite{wang2022systematic}, we enroll the stability mask at our highest temperature range to sample out the most stable bits.
\autoref{fig:stability_bell} shows the distribution of stable bits in 96-bit blocks for both SRAMs. We choose a cutoff of 54, since almost all blocks retain at least that many, giving a leakage in $D_r$ from stability masking of $l_m = 96-54=42$ and $s = 54$ retained stable bits per round.

\begin{figure}[b]
\centering
 \includegraphics[width=0.9\columnwidth]{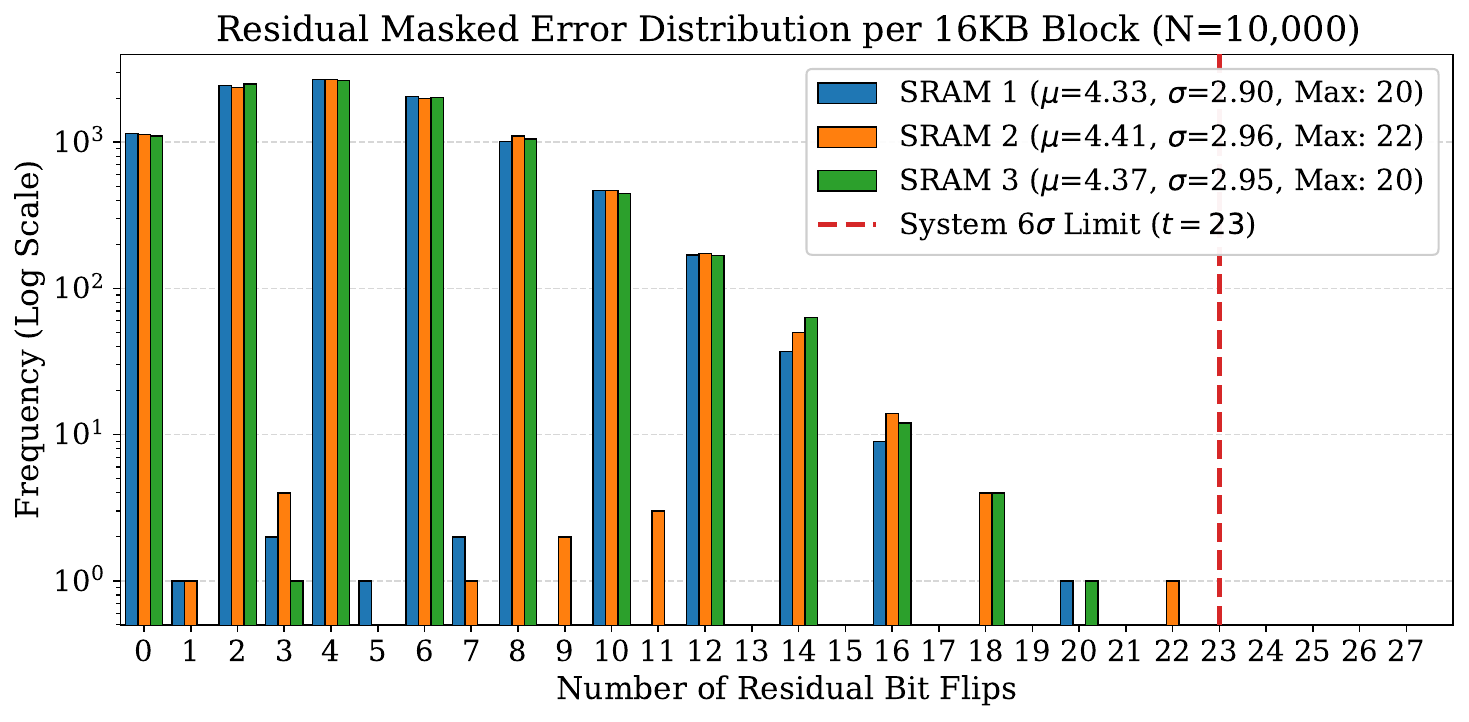}
\caption{Error-distribution of the BCH correction window over 10,000 evaluations for the Cypress CY62256.}
\Description{Graph that shows the error-distribution of the BCH correction window over 10,000 evaluations for the Cypress CY62256. Three differently coloured bars show how many errors each of the three SRAMs had with what frequency.}
\label{fig:bch_bitflip_cy62256}
\end{figure}

\begin{figure}[b]
\centering
 \includegraphics[width=0.9\columnwidth]{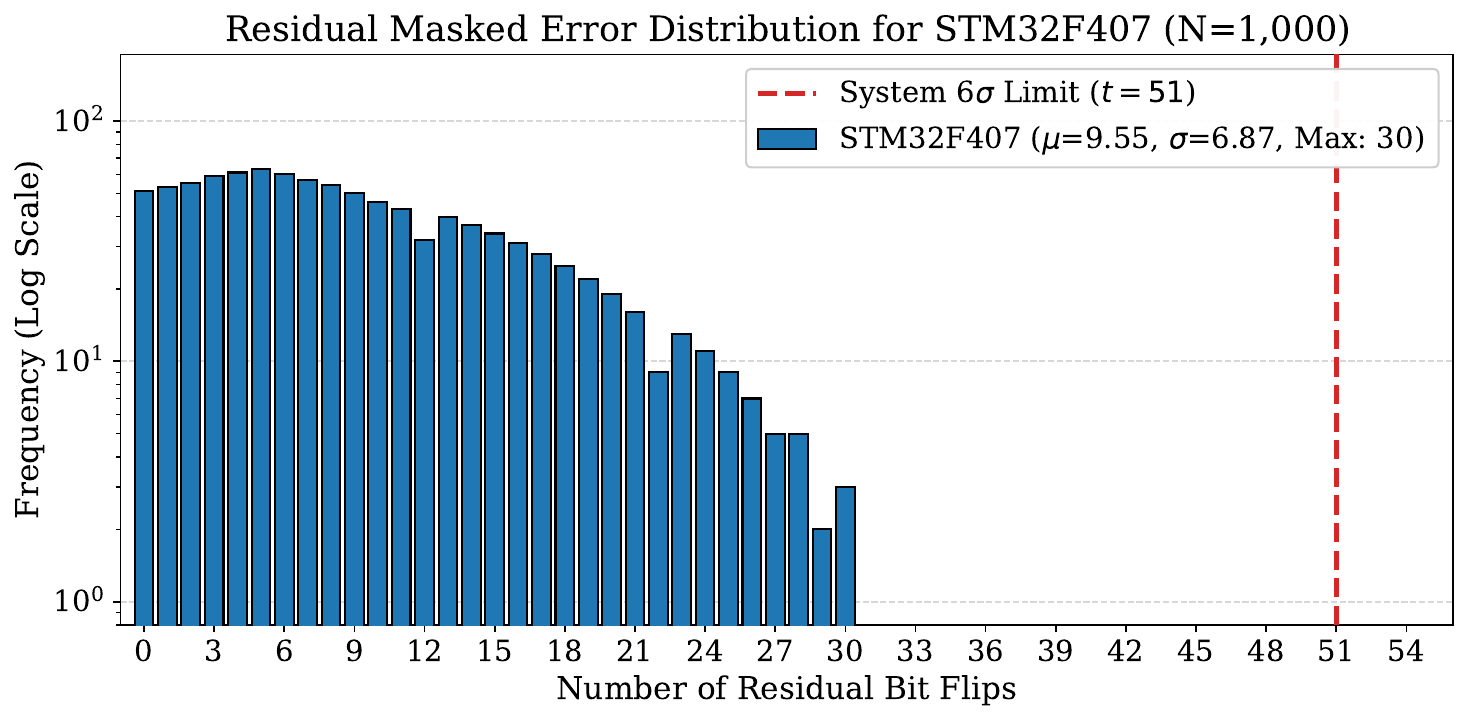}
\caption{Error-distribution of the BCH correction window over 1,000 evaluations for the STM32F407.}
\Description{Graph that shows the error-distribution of the BCH correction window over 1,000 evaluations for the STM32F407. Single coloured bars here since we use just one SRAM split into three blocks.}
\label{fig:bch_bitflip_stm}
\end{figure}

\autoref{fig:bch_bitflip_cy62256} and~\autoref{fig:bch_bitflip_stm} show the distribution of masked errors per $512$-address window, the quantity $t$ has to cover.
We measure at window scale rather than within a single read, since clustering at a larger scale than one read is invisible within-read but shows at the window level.
The Cypress CY62256 figures cover ten thousand power cycles, reported per SRAM since that setup uses three distinct SRAMs. The STM32F407 figure covers one thousand cycles, reported once since its three MSRAM chunks are regions of a single array, and we take fewer cycles since power-cycling manually takes much more time for this board.

The two parts behave differently.
The Cypress chips agree closely with each other, at means $\mu = 4.33, 4.41, 4.37$ and standard-deviations $\sigma = 2.90, 2.96, 2.95$, and we size against the worst of the three.
The STM32F407 is worse on both counts, at $\mu = 9.55$ and $\sigma = 6.87$.
Neither is Poisson.
A Poisson distribution has variance equal to its mean, and a binomial over the window's stable bits would sit slightly below that.
The Cypress runs at a variance-to-mean ratio near $2$ and the STM32F407 near $5$, so flips cluster across addresses on both parts rather than falling independently.
Sizing $t$ from a per-bit rate would therefore have under-corrected substantially, and we use the measured $\sigma$ instead as in~\eqref{eq:t-values}.

We attach confidence intervals to both statistics, since $t$ depends on $\sigma$ six times over and treating a sample estimate as exact would leave the sizing to sampling error.
The mean carries a $95\%$ interval of $\mu \pm 1.96\sigma/\sqrt{N}$ and the standard deviation a chi-square interval on $N-1$ degrees of freedom.
For the Cypress at $N = 10{,}000$ this gives $\mu = 4.41 \pm 0.06$ and $\sigma \in [2.92, 3.00]$, and for the STM32F407 at $N = 1{,}000$, $\mu = 9.55 \pm 0.43$ and $\sigma \in [6.58, 7.18]$.
The smaller STM32F407 sample widens its intervals by roughly a factor of three.
Evaluating~\eqref{eq:t-values} at the upper ends rather than the point estimates leaves $t = 23$ unchanged on the Cypress and raises it from $51$ to $53$ on the STM32F407.
Following this we perform $10,000$ $K_{dev}$ evaluations on the resotred array on Cypress CY62256 and $1,000$ on the STM32F407 and find 0 incorrect evaluations compared to the golden reference.
This test was to check that our BCH variant confirms end to end error-correction of the complete scheme since a correct $K_{dev}$ evaluation confirms the perfect restoration of the array.

Our fourth experiment found that indeed unstable bits are equally likely at every position of the 32-bits read from each SRAM.
This is to check for possible structural leakage in stability masks.

\subsubsection{DUFT}
\label{subsubsec:rtcrresults}
In the previous sections we evaluated $e_{bits}^{CY62256} = 0.38$, $e_{bits}^{STM32F407} = 0.45$, and $l_m^{CY62256} = l_m^{STM32F407} = 42$, giving $s = 54$ retained stable bits per round for both parts.
Both arrays hold $S_{arr} = 442{,}368$ stable bits.
Sizing the window code from~\eqref{eq:nr-conditional} at $win = 512$ addresses with $c = 6$ gives $t = 23$ and $t = 53$ respectively, both over GF($2^{15}$), so $l_{BCH} = 16{,}560$ and $38{,}160$ bits over the three chips.
The deductions are $0.037$ and $0.086$ bits per stable bit, and solving for a target of $n = 128$ bits yields
\begin{align*}
\text{CY62256:} \quad & e_{bits}^{\mathrm{eff}} = 0.343, \; e_D = 18.5, \; n_r = 7
&& \Rightarrow \; 7 \times 54 \times 0.343 = 129.7 \text{ bits} \\
\text{STM32F407:} \quad & e_{bits}^{\mathrm{eff}} = 0.364, \; e_D = 19.7, \; n_r = 7
&& \Rightarrow \; 7 \times 54 \times 0.364 = 137.6 \text{ bits}
\end{align*}
The STM32F407 has the better raw source, at $0.45$ bits per stable cell against $0.38$, but the worse error behaviour, needing $53$ corrections per window against $23$ and paying more than twice the parity for them.
The parity tables occupy $2.0$\,KB and $4.7$\,KB beside the $96$\,KB mask table, and $K_{dev}$ occupies 16 bytes of volatile memory.
The device key adds one BLAKE2s pass over the $96$\,KB restored array per power cycle.
The device key is derived only once per power cycle after the restored array is evaluated so it is not involved in the authentication performance.
Table~\ref{tab:execution_times} provides a breakdown of the performance of the DUFT mechanism when it is used to generate a 128-bit min-entropy response for these SRAMs.
The table includes CPU-time readings for reading the MSRAM, the BLAKE2s hash for the next address, the final BLAKE2s hash, and the complete DUFT mechanism.

\begin{table}[htbp]
    \centering
    \small
    \caption{Execution time comparison for SRAM PUF DUFT (CPU time). Error correction is performed once per power cycle 
    and not on the authentication path.}
    \label{tab:execution_times}
    \begin{tabular}{lrrr}
        \hline
        \textbf{Operation} & \textbf{Invocations} & \textbf{CY62256 ($\mu s$)} & \textbf{STM32F407 ($\mu s$)} \\
        \hline
        Restore Mask and BCH decode  & $1/power~cycle$ & 110000.00 & 6000000.00  \\
        $K_{dev}$ Blake2s  & $1/power~cycle$ & 180.00 & 20000.00  \\
        96-bit Read  & $n_r = 7$ & 700.00 & $<$1.00  \\
        round BLAKE2s & $n_r = 7$ & 20.00 & 30.00  \\
        final BLAKE2s & 1 & 20.00 & 30.00  \\
        \hline
        Complete Algorithm & 1 & $\approx$7000 & $\approx$600 \\
        \hline
    \end{tabular}
\end{table}

Table~\ref{tab:execution_times} shows the constituent operations of DUFT remain in the range of microseconds.
For the Cypress off-chip implementation, the 32-bit read from each SRAM is the most time-consuming operation, since off-chip GPIO transfer is much slower than on-chip embedded SRAM; the STM32F407 performs substantially better on the read path for that reason, though BLAKE2s itself is faster on the Raspberry Pi 5's stronger CPU.
The total execution time of DUFT is then
\begin{equation}
    t_{DUFT} \approx n_r*(t_{SRAM\_Read} + t_{BLAKE2s}) + t_{BLAKE2s} + miscellaneous
    \label{eq:rtcr-time}
\end{equation}
which falls directly with $e_{bits}$, since $n_r$ falls with it.
Restore's masking and decode step runs in milliseconds on the Cypress CY62256 with RPi5/500 setup but takes seconds on the STM32F407. 
The worse performance on the STM32F407 is mainly due to memory. 
The window decoder needs $GF(2^{15})$ log/antilog tables of about 192KB, more than the STM32F407's available SRAM, so they end up in the slower flash instead. 
The random access of these tables mean that cache performance for this is not great either.
The RPi5/500's memory hierarchy does not force this tradeoff. 
Assuming there are a 100,000 authentications per power cycle, the 6s Restore cost adds roughly 60us per authentication, negligible next to the microsecond-scale traversal. 
Regardless, restore happens once per power cycle and does not affect any authentication alone. 
If better performance for restore is desired on either hardware, a smaller BCH window choice would provide it at the cost of more windows and higher parity overhead.

\subsubsection{Protocol}
We next evaluate the complete protocol's performance, measured from NodeA's request to the server through key establishment and authentication.
This is performed only for the Cypress off-chip implementation, due to lack of ethernet support on the STM32F407 board.

\begin{table}[htbp]
    \centering
    \small
    \caption{Execution time comparison for the protocol (CPU time).}
    \label{tab:protocol_times}
    \begin{tabular}{lr}
        \hline
        \textbf{Operation} & \textbf{Execution Time} ($\mu s$) \\
        \hline
        Ascon decryption  & 400.00 \\
        Blake2s HMAC & 10.00 \\
        Packet Delays and Miscellaneous & $\approx$4000.00 \\
        DUFT  & $\approx$7000 \\
        \hline
        Complete Protocol (CPU time) & $\approx$12000  \\
        \hline
    \end{tabular}
\end{table}

As in the standalone mechanism, the off-chip SRAM read dominates the protocol's execution time, and Ascon decryption is small by comparison.
We separate CPU from wall-clock time since packet delays and time blocked on receives are not CPU time, and labelling them as such would misstate the computational cost.

End-to-end performance on the Cypress CY62256 is comparable to other PUF-based protocols, and the single-digit-millisecond performance on the STM32F407 is promising. A better embedded source, with stronger entropy and PUF metrics, would let the round count fall or a stronger min-entropy target be reached.
This performance and entropy estimation, combined with the increased logical challenge space, gives this design utility in several IoT scenarios.

\section{Conclusion}
\label{sec:Conclusion}
In this paper, we addressed the need for extracting responses of a flexible size from SRAM PUFs containing a estimated or established amount of min-entropy, and for supplying a non-repeating stream of challenge-response pairs from a small physical source.
We utilize this improved response in an effective mutual authentication scheme in the IoT space as a lightweight alternative to traditional key-exchange schemes.
A novel mechanism called DUFT was utilized to extract this reponse while providing a large input space to prevent database exhaustion.
The feedback chain is driven by the device's stability mask and device key hashed from its PUF array, which are itself manufacturing fingerprints, so each chip walks a different sequence of cells while keeping the feedback traversal secret.
A physical-source dependent calculation determines the number of rounds and the correction capability needed for a target security level, and thorough error-correction testing was performed such as stability mask profiling and BCH window error statistics.
Some limitations exist.
DUFT provides a large input space, but the physical entropy is still that of the original SRAMs, and no wrapper changes the cost of characterizing that source.
The assumption that the physical adversary has no address-level access to the SRAMs is a strong one.
This is an assumption that is possible to meet and will be required for most weak-PUF based schemes. 
It is met by an embedded SRAM with readout protection enabled and debug disabled.
Our entropy figures come from thirty-two Cypress CY62256 devices and twenty-eight STM32F407 devices and we do not claim they hold across process corners or manufacturing lots.
With these limitations, this work still provides a robust, secure, and lightweight protocol for IoT devices with the minimal resources of SRAMs.
\bibliographystyle{ACM-Reference-Format}
\bibliography{references}

\end{document}